\documentclass[12pt,preprint]{aastex}

\newcommand\ASCA{{\it ASCA}}
\newcommand\BeppoSAX{{\it BeppoSAX}}

\newcommand\HST{{\it HST}}

\newcommand\ROSAT{{\it ROSAT}}
\newcommand\RXTE{{\it RXTE}}
\newcommand\XMM{{\it XMM}}

\newcommand\kms{\ifmmode {\rm km\ s}^{-1} \else km s$^{-1}$\fi}
\newcommand\Hubble{\ifmmode {\rm km\ s}^{-1}\ {\rm Mpc}^{-1} 
	\else km s$^{-1}$ Mpc$^{-1}$\fi}
\newcommand\ctssec{\ifmmode {\rm cts\ s}^{-1} \else cts s$^{-1}$\fi} 
\newcommand\ergsec{\ifmmode {\rm ergs\ s}^{-1} \else 
	ergs s$^{-1}$\fi} 
\newcommand\eflux{\ifmmode {\rm ergs\ s}^{-1}\;{\rm cm}^{-2} \else 
	ergs s$^{-1}$ cm$^{-2}$\fi} 
\newcommand\phflux{\ifmmode {\rm photons\ s}^{-1}\;{\rm cm}^{-2}
	\else  	photons s$^{-1}$ cm$^{-2}$\fi} 
\newcommand\efluxA{\ifmmode {\rm ergs\ s}^{-1}\;{\rm cm}^{-2}\;{\rm
	\AA}^{-1} \else ergs s$^{-1}$ cm$^{-2}$ \AA$^{-1}$\fi} 
\newcommand\efluxHz{\ifmmode {\rm ergs\ s}^{-1}\;{\rm cm}^{-2}\;{\rm
	Hz}^{-1} \else ergs s$^{-1}$ cm$^{-2}$ Hz$^{-1}$\fi} 
\newcommand\cc{\ifmmode {\rm cm}^{-3} \else cm$^{-3}$\fi} 
\newcommand\FWHM{\ifmmode {\rm FWHM} \else ${\rm FWHM}$\fi} 
\newcommand\Msun{\ifmmode M_{\odot} \else $M_{\odot}$\fi}
\newcommand\Lsun{\ifmmode L_{\odot} \else $L_{\odot}$\fi}

\newcommand\Hbeta{\ifmmode {\rm H}\beta \else H$\beta$\fi}
\newcommand\Kalpha{\ifmmode {\rm K}\alpha \else K$\alpha$\fi}
\newcommand\NH{\ifmmode N_{\rm H} \else N$_{\rm H}$\fi}
\newcommand\pasap{{Publ.\ Astron.\ Soc.\ of Australia}}

\shorttitle{Akn 564 \ASCA{} Observations}
\shortauthors{Turner et al.\ 2001}

\begin{document}
	\title{Multiwavelength Monitoring of the Narrow-Line
	Seyfert 1 Galaxy Akn~564.
	I. \mbox{\boldmath $ASC\!A$} 
	Observations and the Variability of the 
	X-ray Spectral Components}
	\author{T.J.\ Turner\altaffilmark{1,2}, 
		P.\ Romano\altaffilmark{3},
		I.M.\ George\altaffilmark{1,2}, 
		R.\ Edelson\altaffilmark{4,5}, 
		S.J.\ Collier\altaffilmark{3}, 
		S.\ Mathur\altaffilmark{3},
		B.M.~Peterson\altaffilmark{3}}

\altaffiltext{1}{Joint Center for Astrophysics,
        Department of Physics, University of Maryland, Baltimore County, 
	1000 Hilltop Circle, Baltimore, 
	MD 21250.} 
\altaffiltext{2}{Laboratory for High Energy Astrophysics, Code 660, 
	NASA/Goddard
  	Space Flight Center, Greenbelt, MD 20771; 
	turner@lucretia.gsfc.nasa.gov, ian.george@gsfc.nasa.gov.} 
\altaffiltext{3}{Department of Astronomy, Ohio State University, 
	140 West 18th Avenue, Columbus, OH  43210;
	stefan@astronomy.ohio-state.edu, 
	mathur@astronomy.ohio-state.edu, 
	peterson@astronomy.ohio-state.edu, promano@astronomy.ohio-state.edu.}
\altaffiltext{4}{Astronomy Department, University of California, Los Angeles, 
                CA 90095-1562; rae@astro.UCLA.EDU}
\altaffiltext{5}{X-Ray Astronomy Group, University of Leicester, 
		Leicester LE1 7RH, U.K.} 

	\begin{abstract}

We present a 35 day  \ASCA{} observation of the Narrow-Line Seyfert 1
galaxy \objectname[MGC +05-53-012]{Akn~564}, yielding an on-source
exposure $\sim$ 1 Ms. The \ASCA{} observation was part of 
a multiwavelength AGN Watch monitoring campaign.
 The soft X-ray light curve binned to 256 s reveals trough-to-peak 
flux variations up to a factor of 
$\sim 16$ and changes in the fractional amplitude of variability across
the observation.

Akn~564 shows small variations in photon index across the 
observation, with 
$\Gamma$ in the range 2.45--2.72. The presence of the soft hump  
component below 1 keV, previously detected in \ASCA{} data, is confirmed.
Time-resolved spectroscopy with $\sim $ daily sampling reveals a 
distinction in the 
variability of the soft hump and power-law components 
over a timescale of  weeks, with the hump varying by 
a factor of $\sim 6$ across the 35-day observation compared 
to a factor 4 in the power-law. 
This difference in the long-term amplitudes of variation 
causes changes in the softness ratio across the observation. 
Flux variations in the power-law component are measured down to 
a timescale of $\sim 1000$ s and accompanying spectral variability suggests the 
soft hump is not well-correlated with the power-law on such short timescales.
However, some correlated events are observed in the soft hump, UV flux  
and hard X-ray flux when all are  sampled daily.
No significant UV to X-ray lags are found, 
with upper limits  $\sim 1$ day.

We detect Fe K$\alpha$ and a blend of Fe K$\beta$ plus Ni K$\alpha$,
the line energies indicating an origin in highly ionized gas. 
Variability measurements constrain the bulk of the Fe K$\alpha$ line 
to originate within $\sim$ a light week of the nucleus. 
The large EW of the emission lines may be due to high metallicity 
in NLS1s, supporting some evolutionary models for AGN.

 	\end{abstract}

	\keywords{galaxies: active -- galaxies: individual (Akn 564) 
	-- galaxies: nuclei -- galaxies: Seyfert -- X-rays: galaxies}

\section{Introduction}	

Narrow-line Seyfert 1 galaxies (NLS1s) are a subclass of active
galactic nuclei (AGN) with relatively narrow permitted optical
emission lines  (\Hbeta{} \FWHM $<$ 2000 $\kms$) 
and a high \ion{Fe}{2}/H$\beta$ ratio \citep{OP85,Goodrich89}. 
Historically, the ratio  [\ion{O}{3}]$\lambda 5007$ $ /$  $\Hbeta < 3$ 
has also been used as a defining criteria for the NLS1 objects, 
although it now appears that this is not a good indicator 
\citep{ra2000}.
Such properties place these galaxies at the lower end of the
line-width distribution for the Seyfert 1 population, thus
distinguishing them from the bulk of Seyfert 1 galaxies (hereafter 
``broad-line Seyfert 1s'' or BLS1s). Furthermore, \citet{BG92} 
find NLS1s to exist as one extreme of their primary eigenvector, 
and conclude that the dominant source of differences in the observed 
optical properties of low-redshift AGN is a fundamental 
parameter (which balances Fe II 
excitation against the illumination of the narrow line region). 

In the X-ray regime, NLS1s exhibit rapid and large-amplitude variability
 \citep[hereafter BBF96]{BBF96}.  \citet{Turnerea99b} showed
X-ray variability to be systematically more extreme in NLS1s than in BLS1s
despite both having the same luminosity distribution. 
For a given luminosity, NLS1s show an order of magnitude greater variability,
as measured by $\sigma^2_{rms}$ \citep{Turnerea99b,Leighly99I}.  
Some NLS1s show X-ray flares 
up to a factor of 100, on timescales of days 
\citep[e.g.,][]{ForsterHalp96,Boll97},  
compared to the factors of $\sim$ a few seen in BLS1s. 
 \ROSAT{} (0.1--2.4 keV)
observations revealed that the NLS1 soft X-ray continuum slopes are
systematically steeper than those of BLS1s (BBF96), the photon index $\Gamma$
(photon flux $P_{\rm E} \propto E^{-\Gamma}$) sometimes  exceeding 3. This
phenomenon was found to extend  to higher energies using  \ASCA{} (2--10 keV)
observations \citep{BME97,TGN98,Leighly99II,Vaughan99b}. The very strong
anticorrelation between \Hbeta{} \FWHM{} and both the X-ray spectral slope in Seyferts 
(BBF96) and in quasars \citep{Laorea97} and ``excess variance''  
\citep{Turnerea99b}
suggests that the remarkable X-ray properties of NLS1s represent an extreme of
Seyfert activity, possibly due to an extreme value of a fundamental physical
parameter related to the accretion process. 

A popular explanation of the differences in X-ray 
properties across the Seyfert population is that 
NLS1s have relatively low masses for the central black hole. 
Smaller black-hole masses naturally explain both the narrowness of
the optical emission lines, which are generated in gas that has
relatively small Keplerian velocities, and the extreme
X-ray variability, since the primary emission would originate in a
smaller region around the central engine \citep[e.g.,][]{Laorea97}.
Given that NLS1s have comparable
luminosity to that of the BLS1s, \citet{PDO95} suggested that 
they must be emitting at higher 
fractions of their Eddington luminosity, hence higher fractional
accretion rates ($\dot{m} = \dot{M}/\dot{M}_{\mbox{\scriptsize Edd}}$) 
are also required. 

The closer the luminosity to the
Eddington limit (and the lower the black-hole mass), the greater the
fraction of the energy emitted by the accretion disk in the soft
X-rays \citep{RFM92}. Thus NLS1s might be expected to show 
disk components which peak at higher energies than for BLS1s. 
\citet{PDO95} and \citet{MH97} noted that soft photons from the disk may 
Compton-cool hard X-rays from the corona, and 
cause the steep observed photon indices. 
In the case of a high accretion rate, the surface of the disk is 
also expected to be highly ionized \citep{MFR93}; evidence 
of reflection 
from such a disk (from Fe K$\alpha$) is found in six NLS1s: 
\citet{Comastriea98}, \citet{TGN98} for Ton S180;
\citet{Vaughan99a}, \citet{Comastriea00}, \citet{TGN99} for Akn 564;
\citet{BIF00} for the previous sources plus Mrk 335, NGC 4051, PG
1244+026, and PKS 0558-504. 
However, lines from ionized Fe are not unique to the NLS1 population, 
since they sometimes occur in BLS1s as well \citep[e.g.,][]{Guainazziea98}.

Alternative explanations of the difference in observed parameters 
across the Seyfert population include the possibility that NLS1s have 
broad-line regions more distant from the nucleus 
than BLS1s and hence smaller Keplerian 
line widths \citep{GFM83}, or that NLS1s are observed preferentially close 
to face on. 
The latter is disfavored by the analysis of \citet{BG92}, and the fact 
that the inner regions of BLS1s also appear to be observed close 
to face-on \citep{Nandraea97a}. 
The former possibility can be tested by measuring the size of the 
broad-line region using reverberation mapping. 

Reverberation mapping uses measurements of 
the correlations between emission-line and continuum 
flux variations across a broad energy band  
to determine the size-scales of the emitting regions  \citep{P93}. 
Line profiles give the kinematics of the emitting gas, and 
combining the size and velocity information yields an estimate of 
the central mass of an AGN \citep{wpm99}. 
Results obtained to date have indicated that for NLS1s 
 $M_{\rm BH} \approx 10^{6} \Msun$, as opposed to 
$M_{\rm BH} \approx 10^7 - 10^8\,\Msun$ 
for BLS1s, while NLS1s and BLS1s appear to have broad-line regions 
which are comparable in size \citep{K2000,P2000}. 

Akn~564 (IRAS 22403+2927, MGC +05-53-012) is a bright, nearby 
NLS1  with $z = 0.0247$, \citep{Huchraea99}; $V = 14.6$ mag, 
\citep{rc3.9catalogue}, and 
$L_{\mbox{\scriptsize 2--10{} keV}} \approx 2.4 \times 10^{43}$
\ergsec (from these data);  hence an ideal test case for accretion
disk models of AGN. 
\ROSAT{} observations 
\citep{BFNRB94}  showed a complex soft X-ray spectrum with
spectral features around $\sim 1$ keV. However, the {\it
PSPC} data could not discriminate between absorption or emission 
as the origin of the complex form of the spectrum. 
Subsequent \ASCA{} observations ($\sim 50$ ks, 
\citealt{TGN99,Vaughan99a,Leighly99II}) confirmed
the complexity of the soft spectrum and showed a strong iron line at
rest energy of about 7 keV, indicative of reflection from highly
ionized material. 
This conclusion was supported by 	%
\citet{Vaughan99b} who detected an edge at 8.5 keV 
in combined \ASCA{} and \RXTE{} data, attributable to 
reflection from a
strongly irradiated disk. \BeppoSAX{} data \citep{Comastriea00} are 
also indicative of reflection from a highly ionized, optically-thick
accretion disk.

The most intensive broad-band reverberation mapping program 
undertaken to date aimed to determine the nature of the 
relationship between X-ray and UV--optical continuum variations 
and thus obtain an estimate of the BLR size and
virial mass of the central source in the NLS1 galaxy, Akn~564. 
This campaign was designed to obtain the first estimate for a NLS1 
of the size of the region emitting the broad UV lines, and it
provided the longest baseline for a quasi-continuous  X-ray study of 
any AGN to date (i.e interrupted only by earth occultation, SAA passage 
etc). 
Akn~564 was observed by \ASCA{} on 2000 June 1 to July 6, 
during a multiwavelength  monitoring campaign that included
observations from  
\HST{} between May 9 and July 8 \citep[][Paper II]{Collierea01},
\RXTE{} between June 1 and July 1 \citep[][]{Poundsea01,Edelsonea01}, 
and \XMM{} \citep[][]{Brandtea01} June 17,
and from many ground-based observatories as part of the 
AGN Watch\footnote{\anchor{http://www.astronomy.ohio-state.edu/~agnwatch}
{All publicly available data and complete references to published 
AGN Watch papers can be found at 
http://www.astronomy.ohio-state.edu/$\sim$agnwatch.}} 
project \citep[][Paper III]{Shemmerea01}. 

In this paper we present the results from the  
$\sim$ 1 Ms ASCA observation of Akn~564.  
In \S~\ref{dataobs} we present the data; in \S~\ref{timevar} 
we 
discuss the time variability of the source; 
\S~\ref{meansp}
we analyze the mean spectrum; in \S~\ref{TimeSel} we discuss 
time-resolved spectroscopy; 
\S~\ref{rms} presents an investigation of the RMS spectrum;  
\S~\ref{CCresults} looks at the correlations between the X-ray 
parameters, and with the UV flux at 1365 \AA ; 
\S~\ref{results} presents a summary of our observational 
results; finally, in \S~\ref{discussion} we discuss 
the X-ray results.

	\section{Observations and Data Reduction
			\label{dataobs}} 

\ASCA{} is equipped with four focal-plane detectors that are operated
simultaneously, namely, two CCDs (the Solid-state Imaging
Spectrometers SIS-0 and SIS-1, 0.4--10 keV, \citealt{Burkeea91})
and two gas-scintillation proportional-counters (Gas Imaging
Spectrometers  GIS-2 and GIS-3, 0.7--10 keV,
\citealt[][and references therein]{Ohashiea96}). 
ASCA obtained a total duration of $\sim 2.98$ Ms on Akn~564, 
starting from JD $= 2451697.024$ (for the screened data).
The data were reduced using standard techiques as described in
\citet{Nandraea97a}; in particular, the methods and screening
criteria utilized by the {\it Tartarus}\footnote{
\anchor{http://tartarus.gsfc.nasa.gov}{http://tartarus.gsfc.nasa.gov}} 
database \citep{Turnerea99b} were used.
Data screening yielded an effective exposure time of $\sim 1.11$ Ms
for the SISs and $\sim 1.29$ Ms for the GISs. The mean SIS-0 count
rate was 1.894 $\pm 0.001$ \ctssec{} (0.6--10 keV band).

\label{SIStrouble}
Since launch in 1993, the \ASCA{} SIS detectors have been
experiencing a degradation in efficiency at the lower energies, which
is probably due to increased dark current levels and decreased 
charge transfer efficiency (CTE), producing SIS spectra which
diverge from each other and from the GIS data. This loss in
efficiency is not well understood and therefore not 
corrected for by any of the software (in particular,
{\tt CORRECTRDD} does not solve the problem), so that the discrepancy
between SIS-1 and GISs can be as much as 40\% below 0.6 keV for data
taken in January 2000\footnote{ 
\anchor{http://heasarc.gsfc.nasa.gov/docs/asca/watchout.html}
{see http://heasarc.gsfc.nasa.gov/docs/asca/watchout.html}}.
Data from the last phase of {\it ASCA} operations, AO-8, 
revealed that there has been a non-linear evolution of the SIS 
CTE. The Akn~564 data were 
first calibrated using linear extrapolations of SIS gain from 
the last determination of CTE values on 1997 March 11. That 
calibration produced unacceptably large inconsistencies between the 
SIS and GIS instruments up to a few keV. Thus 
the previous solution of excluding SIS data below 0.6 keV 
is inadequate. We recalculated the SIS gain using the interim 
solution released on 2001 February 13 (using CTE file 
$sisph2pi\_130201.fits$). This interim solution reduced the 
instrument discrepancies significantly, 
but still necessitated the exclusion of 
some sections of the SIS data. For this reason, we 
proceed as follows. We have 
modelled the features which appear in both GIS and SIS detectors 
and ignored residuals which are in one instrument or instrument pair only 
(and are likely artifacts of calibration problems). We then 
used our simple parameterization of the data to 
examine the time-variability of the source spectrum. 

\label{Yaqoob}
The divergence of the SIS detectors at 
low energies can be partly corrected for in the spectral analysis. 
One approach \citep{Yaqoob}\footnote{
\anchor{http://lheawww.gsfc.nasa.gov/$\sim$yaqoob/ccd/nhparam.html}
{see http://lheawww.gsfc.nasa.gov/$\sim$yaqoob/ccd/nhparam.html}}, 
is to quantify the apparent loss in SIS efficiency as a function of
time throughout the mission; an empirical correction can be applied
by parameterizing  the efficiency loss with a time-dependent 
absorption (``excess \NH''). The correction for SIS-0 follows the linear
relationship, $\NH(\mbox{SIS0}) = 3.635857508 \times 10^{-8}\;  (\mbox{
{\tt T}}
-3.0174828 \times 10^{7}) \; 10^{20}$ cm$^{-2}$, where {\tt T} is the 
average of start and stop times of the observation, based on 
seconds since launch. There
is no simple analytical relationship that satisfies the SIS-1 excess
\NH, but it is usually found that 
a slightly larger column can be applied to the SIS1 data to bring it
into line with SIS0. 
Naturally, one has to be extremely
conservative in interpreting spectral features derived from fits in
the lower-energy region, in particular  below 1 keV, once this
empirical correction has been applied. For our observation, where
{\tt T} $= 2.36 \times 10^{8}$ s, $\NH(\mbox{SIS0}) =7.5 \times
10^{20}$ cm$^{-2}$ and we adopted $\NH(\mbox{SIS1}) = 1.05 \times
10^{21}$ cm$^{-2}$.

A more conservative approach consists of only considering data above
1 keV for both SIS and GIS, where the disagreement is less 
dramatic. However, as we are very interested in 
the variability of the soft hump spectral component, in this paper, 
we will use the excess $\NH{}$ correction for our fitting.

	\section{The Time Variability\label{timevar}}

\begin{figure}	
        \epsscale{0.85}
        \plotone{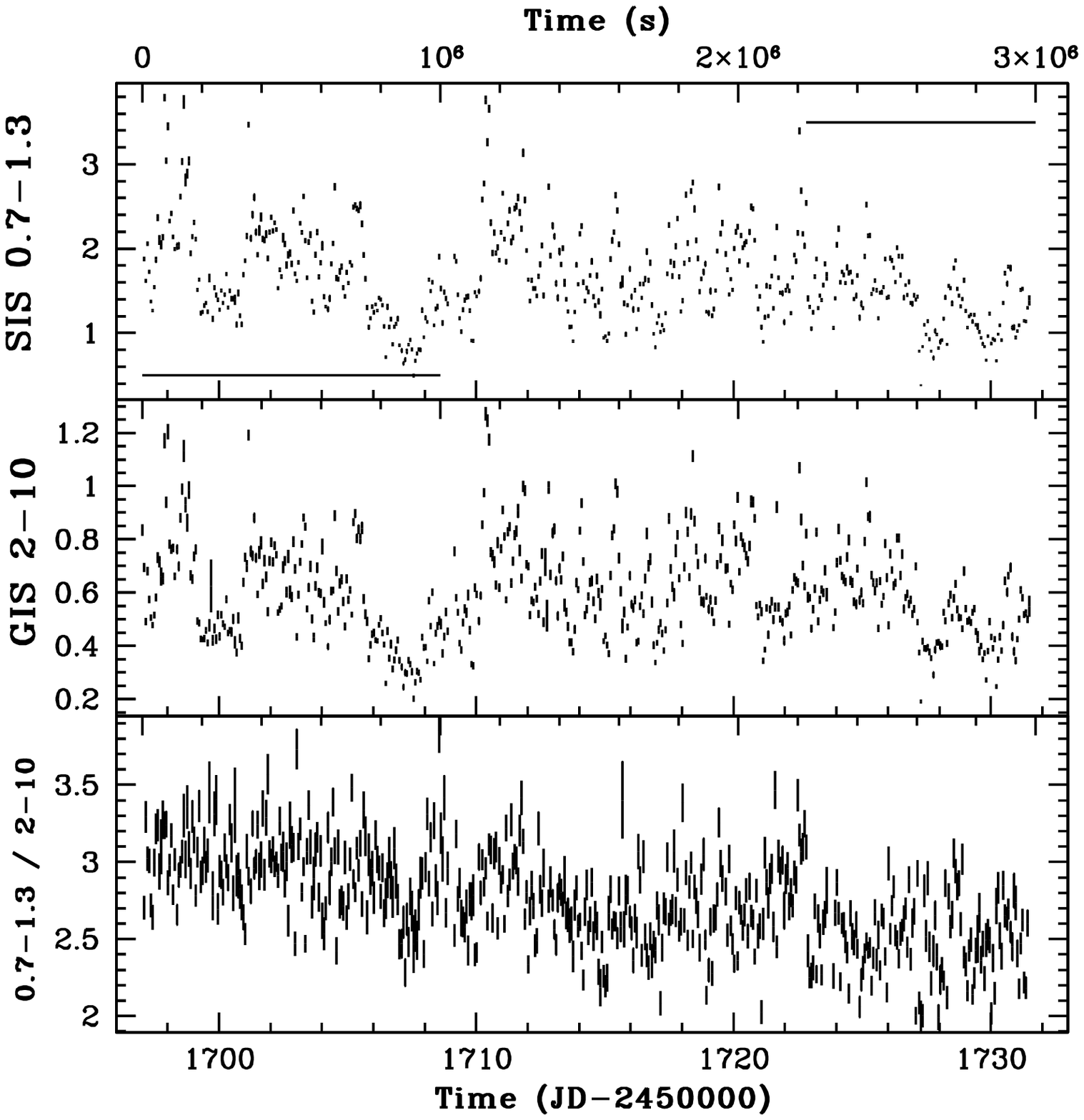}
        \caption[Light curves]{Light curves 
for the \ASCA{} data taken between
2000 June 1 and July 6, in \ctssec{} and in 5760 s bins. 
The top panel is the SIS soft band (0.7--1.3 keV) 
light curve; the middle panel the GIS hard band (2--10 keV) 
and the bottom panel is the ratio of 0.7--1.3/2--10 keV. 
The background level in the source cell is about 4 \% of the source 
count rate, and not plotted. The times are reported both in seconds
from the start of exposure (top axis) and in JD-2450000 (bottom axis).
The horizontal lines show the periods referred to as the soft and hard 
states. 
\label{lcv2}}
\end{figure}

Light curves were extracted using bin sizes of 256 s, and
5760 s  in the full-band (0.7-10 keV)  for the SIS, 
a soft-band (0.7--1.3 keV) for the  SIS data, 
and the hard-band (2--10 keV) 
for both GIS and SIS data. The use of 0.7 keV as a lower limit 
for the SIS data was due to the high setting for the 
SIS lower level discriminator towards the end of the observation. 
In all cases we 
 combined the data from the SIS and GIS detector pairs.
The exposure requirements for the combined curves were that
the bins be fully exposed in each instrument for the 256 s curves 
and at least 10 \% exposed for the 5760s curves.  
The observed count rates correspond to a mean 2--10 keV flux of
$2 \times 10^{-11}$ ${\rm erg\ cm^{-2}\ s^{-1}}$ and 
2--10 keV luminosity $2.4 \times 10^{43}$ \ergsec{}  
(assuming $H_0=75$ ${\rm km\ s ^{-1}\ Mpc^{-1},}$  $q_0=0.5$). 
This mean flux level is $\sim$ 20 \% brighter than 
that observed during a previous \ASCA{} observation 
in 1996 December 23  \citep{TGN99}.  
Figure~\ref{lcv2} shows the combined 0.7--1.3 keV SIS soft-band and 
GIS hard-band light curves in 5760 s bins.
The background level in the source cell is about 4 \% of the source 
count rate, and not plotted or subtracted. 
Figure~\ref{lcv2} also shows the softness ratio, defined as the ratio  
between the count rates in the 0.7--1.3 and 2--10 keV bands.  
The spectrum hardens during the \ASCA{} observation.  

The light curves binned to 5760 s show trough-to-peak variations 
in flux by a factor of $\sim 10$ in the 0.7--1.3 keV band (SIS), 
$\sim 7$ in the 2--10 keV band (GIS). 
Examination of light curves binned to 256 s reveals even larger 
amplitudes due to fast flickering. The maximum amplitude 
of variability in these (256 s) curves is a factor $\sim 16$ for 
the SIS data in the 0.7-1.3 keV band and $\sim 14$ for the GIS 
data in the 2--10 keV band. 
In their Figure~3, \citet{Edelsonea01} show the 0.7--1.3 keV 
and 2--10 keV light curves, 
in 256 s bins, for a ``flare''  at JD $\approx$ 2451710 
(centered around 1153 ks after the start of the observation). 
The SIS and GIS data show rate variations of a factor of 1.4 and 1.8 
(respectively) in 
$\Delta t =1280$ s, corresponding to a variation in luminosity of 
$\Delta L = 1.9 \times 10^{43}$ \ergsec. The close-up examination 
of that flare shows 
that the hard X-ray event is sharper than that in the soft band, 
and that there is a significant 
change in softness ratio on a timescale of $\sim 1000$ s 
\citep{Edelsonea01}. 
Close examination of the light curve reveals 
other similar examples of rapid changes in spectral shape 
and also some flares which are not accompanied by 
a strong spectral change. The behavior of the source is obviously 
complex with some events characteristically different to others. 

Here we concentrate on the search for variations in spectral 
parameters across the 35 day baseline of the observation 
to understand the nature of the spectral variability. A 
more detailed timing analysis is presented in a complementary 
paper, \citet{Edelsonea01}. 

	\subsection{Fractional Variability Amplitude\label{sigmavar}}

The fractional variability amplitude $F_{\rm var}$ is defined 
in \citet{Edelsonea01}, as 
\begin{equation}
F_{\rm var} = \sqrt{\frac{S^2 - \langle \sigma^2_{\rm err} 
\rangle}{\langle X \rangle^2} } ,
\end{equation}
where $S^2$ is the total variance of the light curve, 
$\sigma^2_{\rm err}$ is the mean error squared, and 
$\langle X \rangle$ is the mean count rate. 
The error on $F_{\rm var}$, also from \citet{Edelsonea01}, is 
\begin{equation}
\sigma_{F_{\rm var}} = \frac{1}{F_{\rm var}} 
		\sqrt{\frac{1}{2N}} \frac{S^2}{\langle X \rangle^2} .
\end{equation}

The quantity $F_{\rm var}$ was calculated every $\sim$ day 
using even sampling and constructing light curves with 256 s bins. 
We examined the quantity in the soft (0.7--1.3 keV) band and the 
hard (2--10 keV) band. These bands were chosen to be as widely separated as 
possible in the energy-bandpass, but having good signal-to-noise 
in each curve. $F_{\rm var}$ is variable across the observation, 
but there appears to be 
no correlation between this quantity and the flux-state of the source, 
probably due to a random nature of the light curve. 
$F_{\rm var}$ appears very well correlated between the soft and 
hard bands. 
As we will demonstrate in \S~\ref{meansp},  
in the mean spectrum the power-law  continuum 
provides about 75 \% of the flux in the 0.7--1.3 keV band. The 
rapid variations in the power-law  component dominate the variability  
so the two bands have a correlated component of variability, making 
$F_{\rm var}$ appear similar, to first order, 
 in the soft and hard bands.

$F_{\rm var}$ calculated across the whole dataset is 
34.84 \% $\pm$ 0.46 for the 0.7--1.3 keV band, and 
33.19 \% $\pm$ 0.40 for the 
2--10 keV band. This quantity measures deviations 
compared to the mean, integrated over the entire month. The 
slightly greater 
$F_{\rm var}$ for the soft band is probably due to the relatively strong 
variation in the soft component occurring over a month timescale, 
as we will show in \S~\ref{humpfits}.

	\section{The Mean Spectrum\label{meansp}}

For the spectral analysis the source counts were grouped with a
minimum of 20 counts per energy bin. After examining the spectral 
fits separately to quantify the cross-calibration problems 
(as described in \S~\ref{SIStrouble}), 
the data were fit allowing the relative normalization of the four 
instruments to be free to allow for small differences in calibration 
of the absolute flux, and differences in the fraction of encircled counts 
contained within the SIS and GIS extraction cells. 
The spectral fits have been performed with the {\tt XSPEC V11.0.1} 
package, using the response matrices released in 1997 for the GIS, 
and response files generated using {\tt HEAsoft v5.0.4} for the SIS. 

The spectral continuum slope was first determined by fitting a power-law 
model modified by Galactic absorption 
($\NH = 6.4 \times 10^{20}$ cm$^{-2}$, \citealt{DickeyL90}), 
and corrected for the SIS low-energy problem as described 
in \S~\ref{SIStrouble}. 
For this fit we used data in the bandpass 
2--5 plus 7.5--10 keV data in the rest-frame ($\sim$ 1.8--4.9 keV 
observer's frame). Furthermore, we excluded SIS data 
in the 1.7--2.5 keV regime and above 7.32 keV (both observer's frame) 
due to problems in the calibration that show up in data of such high 
signal-to-noise. 
The power-law fit yielded $\Gamma = 2.538 \pm 0.005$ and 
$\chi^2=1403$ for 1225 degrees of freedom ($dof$). 
This and subsequent errors represent, unless otherwise specified, 
the 90 \% confidence level. 

\begin{figure}	
	\epsscale{0.85}
	\plotone{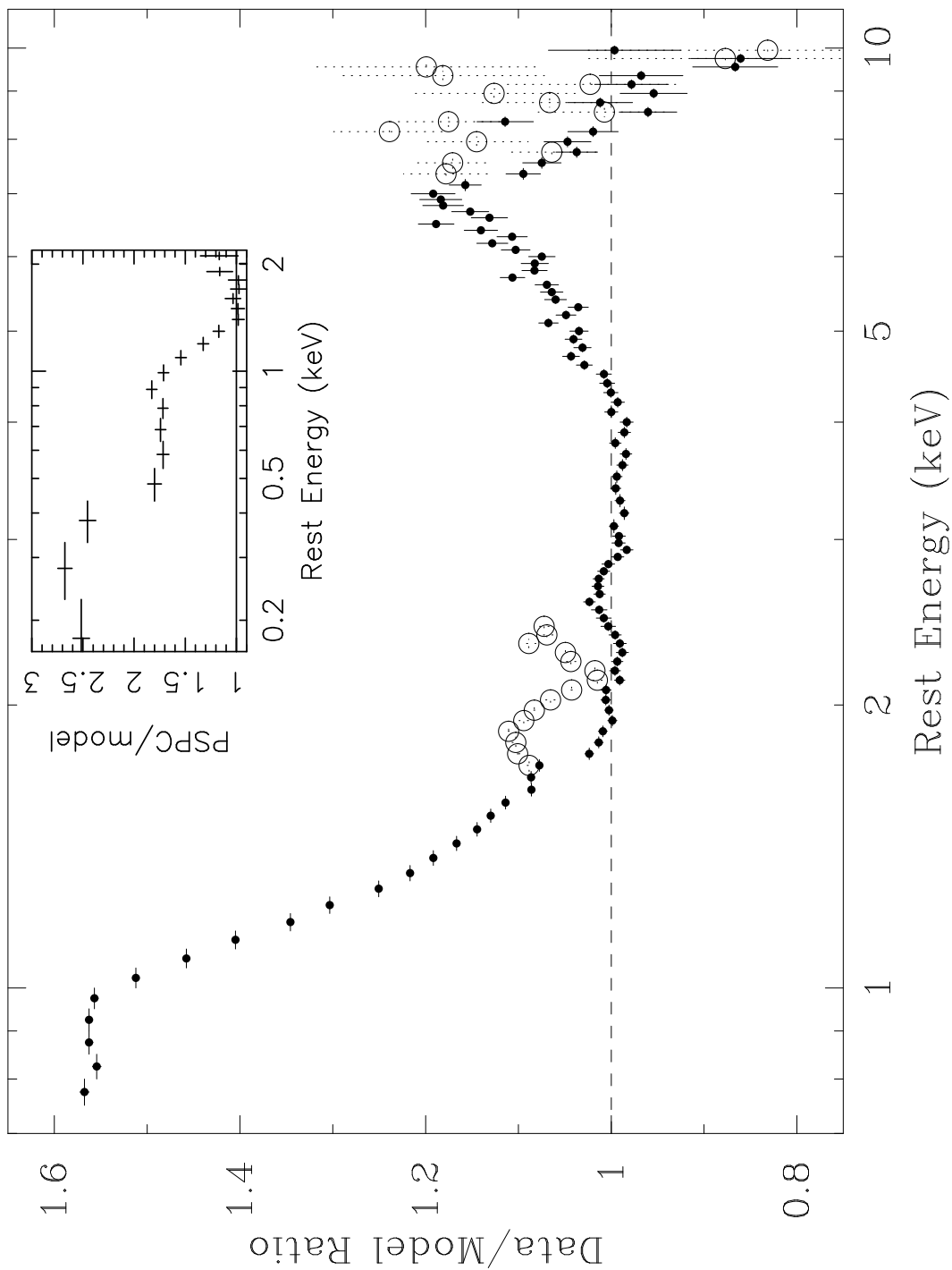}
	\caption[Data/Model Ratio]{Data/Model where the model 
is a simple power law fit to the 2--5 plus 7.5--10 keV data 
(rest-frame). The rest of the ASCA data are then overlaid, 
revealing a strong soft hump and Fe emission line.
The data used in our spectral analysis are shown as filled 
circles, the data disguarded due to problems with the current 
ASCA calibration are overlaid, as open circles. The inset panel 
shows the {\it ROSAT} PSPC data compared to the continuum power-law,
demonstrating the shape of the soft hump to lower energies. 
\label{pl_rat}}
\end{figure}

The data/model ratio is shown in Figure~\ref{pl_rat}, along with the 
good data overlaid relative to this continuum model, 
and also the bad (excluded) data, shown as a different  point-style.
A strong soft excess is evident which appears as a hump of emission, 
rising above the power-law continuum below 2 keV, then flattening 
off below 1 keV, as observed in an earlier \ASCA{} observation 
\citep{TGN99}; hereafter we refer to this component as the soft hump. 
It is also interesting to note that the shape of the soft hump 
is evident in the PSPC spectrum. We constructed a 
data/model spectral ratio for the archival {\it ROSAT} 
PSPC data from 1993. The data were compared to a power-law 
model of the mean slope noted above. The PSPC data  
between 1.5--2.0 keV were used to normalize this component 
(there being insufficient bandpass to determine the hard X-ray slope) 
and the rest of the spectral data were then overlaid. The ratio 
is shown in the inset panel of Figure~\ref{pl_rat}, 
demonstrating that the soft X-ray 
spectrum turns up again between 0.4--0.5 keV. (Even if the power-law 
index is different at the epoch of the PSPC observation, we could 
not introduce the structure observed in the soft X-ray spectral shape.)
Also evident in the main panel is 
an excess of emission close to 7 keV, which we know to be 
due to an unmodeled Fe K$\alpha$ line.
The overlay of the excluded bad points serves to show where 
the strongest problems are, due to 
the detector aging and calibration issues 
described in \S~\ref{SIStrouble}.

\subsection{The Soft Component}
\label{meanhump}

The status of the calibration, the degradation 
of the SIS energy-resolution towards the end of the mission,  
and the small bandpass of data available to examine the soft hump component 
conspire to make it impossible to achieve an unambiguous parameterization 
of the hump shape using these data. This led us to 
use a very simple parameterization of the hump in order 
to simply examine its flux variability. 

A previous detailed study of Akn~564 \citep{TGN99} 
detected the soft hump and  ruled out origins 
solely due to 
the effects of emission and/or absorption from photoionized gas. 
Models based upon emission from thermal gas were
consistent with the spectral data, but posed problems in terms 
of physical consistency with the picture of an AGN \citep{TGN99}.  
In light of the 
{\it Chandra} results for the NLS1s Ton S180 \citep{Turnerea01} 
and NGC~4051 \citep{Collea01} 
that the hump is apparently 
a smooth continuum component,  
a variability study such as this has turned out to be a better 
technique to understand the nature of the soft hump 
than high resolution spectroscopy, as shown below in \S~\ref{humpfits}. 

The lack of constraints on the form of the soft hump  
lead us to use a  simple Gaussian 
parameterization of the component. This adequately  
models the shape and flux of the 
excess in the \ASCA{} data, and allows us to study the variability 
of the component using time-resolved spectroscopy. 
Using SIS data in the range 0.75--1.7 and 2.5--4.88 keV 
simultaneously with GIS data in the range 1.0--4.88 and  
7.32--9.76 keV (observer's frame), the Gaussian model which best 
fits the excess has 
a peak energy $E=0.57\pm0.02$ keV, width $\sigma=0.36\pm0.01$ keV 
and normalization $n=1.25^{+0.12}_{-0.17} \times 10^{-2}$  
$\phflux$ corresponding to a mean equivalent width (EW)$=110^{+11}_{-15}$
eV.

\subsection{The Fe K$\alpha$ Regime}
\label{meanfek}

A significant Fe K$\alpha$ emission line is evident in the 
{\it ASCA} spectrum (Figure~\ref{pl_rat}). The 
line profile is asymmetric with a marked red wing, as 
observed in Seyfert 1 galaxies  \citep{Nandraea97b},  
but with a peak close to 7 keV. These basic properties of the line 
will be robust to refinements to the instrument calibrations. 

We utilized SIS data in the range 2.5--7.32 keV 
simultaneously with the GIS data in the range 1.8--9.76 
keV. A narrow Gaussian component is an inadequate model for 
the line profile, 
but allows us to determine that the line peak is at an observed 
energy of 7.1 keV. 
The asymmetry of the line led us to fit the data with a 
disk-line model profile \citep{Fabianea89}. The model 
assumes a Schwarzschild geometry and we assumed an emissivity law 
$r^{-q}$ for the illumination pattern of the accretion disk, where 
$r$ is the radial distance from the black hole.
We assume the line originates between 6 and 1000  
gravitational radii ($R_{\rm g}$) 
and we constrained the rest-energy of the line to lie between 6.4 and 7 keV 
(as this first test is for Fe K$\alpha$ and this range represents 
the rest-energies possible for this line,  depending on ionization-state 
of the reprocessing gas). 
The inclination of the system is defined such that $i=0$ is a disk 
oriented face-on to the observer. 
This model  gave 
$\chi^2=6584$ for 2481 $dof$ for a fit including the full
range of good data. The rest-energy of the line
was $E = 7.00^{+0.00p}_{-0.13}$ keV, i.e. the energy pegged at the 
upper limit allowed in the fit. 
The inclination was $i=26\pm2$ degrees, emissivity 
index was $q=5.7\pm0.9$ and normalization 
$n=4.80\pm0.50 \times 10^{-5}$ \phflux.
The equivalent width was EW$=351^{+29}_{-37}$ eV. 
The index was $\Gamma=2.541^{+0.006}_{-0.004}$ in this model, 
i.e. consistent with that obtained by fitting for the continuum alone. 
While the fit is statistically poor, the contributions to 
$\chi^2$ are dominated by a few areas 
where the data from the four instruments  
diverge, due to problems with the calibration for data from this epoch. 
No systematic residuals remain which are present in all instruments.  

Next we tested a model for the line profile  assuming a Kerr metric, 
as implemented by \citet{Laor91}, for a maximally rotating 
black hole which will have the most intense gravitational effects.
In this case we fixed the innermost radius as the last stable orbit 
for a Kerr hole, and the outer radius at the maximum value allowed 
by the model, 
400 $R_{\rm g}$. 
The Kerr model provides a fit-statistic  $\chi^2=6573$ for $2481\ dof$, 
an improvement (at 95 \% confidence) over the Schwarzschild model. 
The rest-energy of the line 
was $E = 6.99^{+0.01p}_{-0.13}$ keV, inclination was 
$i=17^{+11}_{-17}$ degrees, emissivity index was $q=3.25\pm0.14$
and normalization
$n=8.04\pm0.50 \times 10^{-5}$ \phflux.
The equivalent width was EW$=653\pm85$ eV. The photon index was 
$\Gamma=2.583^{+0.019}_{-0.003}$. 

Analysis of previous short \ASCA{} observations revealed some ambiguity 
as to whether the Fe K$\alpha$ line arose from ionized material, or a disk 
highly inclined to the line-of-sight \citep{TGN98}, 
therefore we explicitly tested for an origin in neutral material. 
If the line energy is fixed at 6.4 keV, but all other parameters 
are allowed to vary, 
then the fit-statistics are $\chi^2_{\rm Schwarz} =6653/2482\ dof$ 
for the Schwarzschild model, and $\chi^2_{\rm Kerr} =6607/2482\ dof$
for the Kerr model. These fits are significantly worse than 
those which show a line energy close to 7 keV, thus we conclude that 
the Fe K$\alpha$ (Ly$\alpha$) 
emission in Akn~564 does originate in highly ionized material, 
dominated by emission from Fe{\sc xxvi} (H-like) ions. 
The line energies derived are indicative that H-like Fe dominates more
than He-like Fe. However, we consider this to be an area which 
requires revisiting with the forthcoming improvements to
GIS and SIS  calibration, and drawing firm conclusions on the 
relative importance of He-like and H-like ions 
would be premature with the calibration used here. 

\begin{figure}	
	\epsscale{0.85}
	\plotone{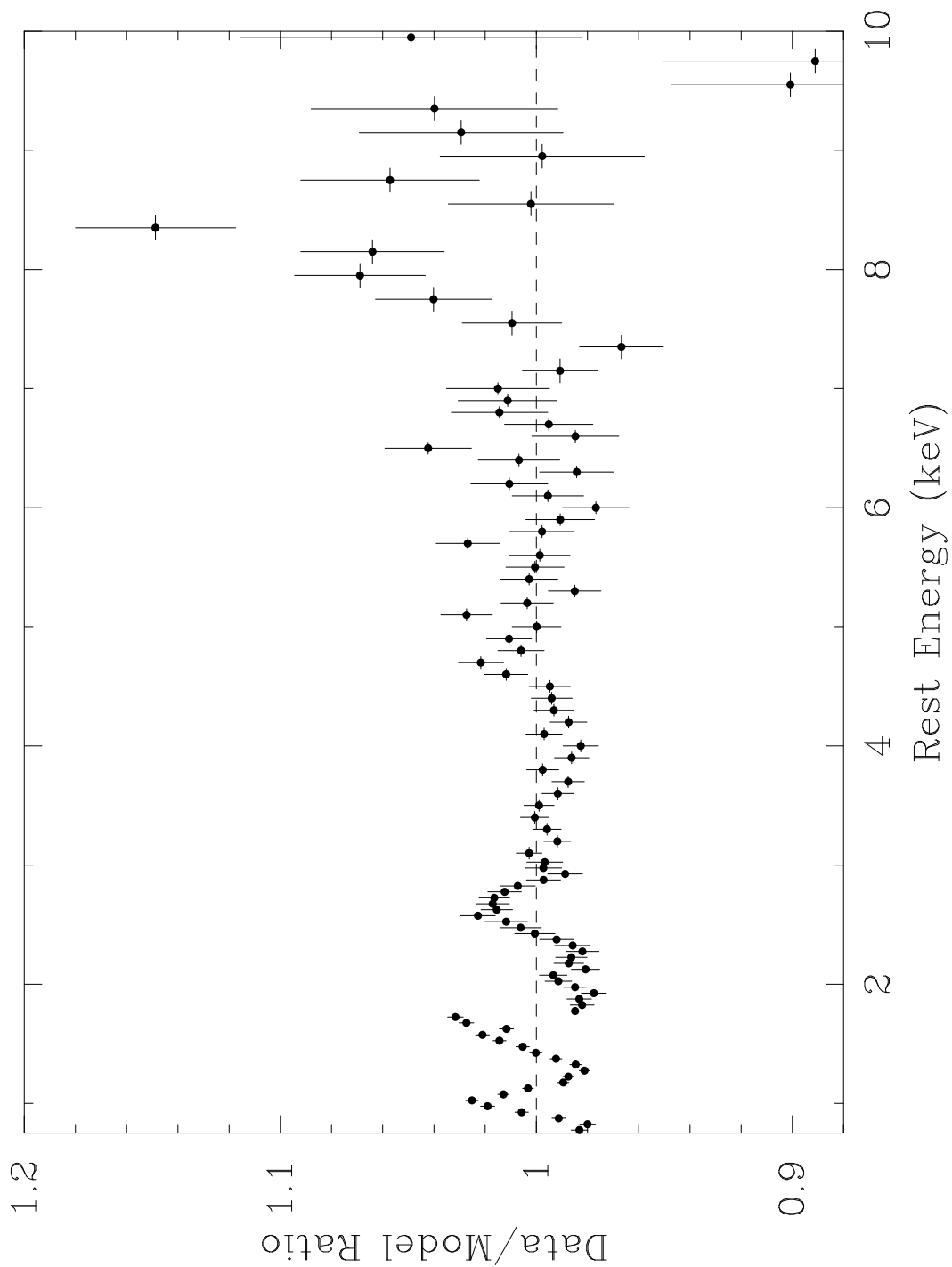}
	\caption[Data/Model Ratio]{Data/Model where the model 
is a power law plus soft hump plus 
Kerr model for the Fe K$\alpha$ line. The 
data show the presence of an additional component, due to a blend 
of Fe K$\beta$ and Ni K$\alpha$ emission at a rest-energy $\sim$ 8.2 keV. 
\label{lybeta}}
\end{figure}

The residuals to the Kerr fit are shown in Figure~\ref{lybeta}.
There is a significant excess of emission suggestive of 
an additional emission line just above 8 keV (this shows up in both 
the GISs whose data we used, and examination of the disguarded SIS data
in this range also revealed the line component).
 Using a narrow Gaussian 
model to fit this excess revealed a line energy 
$E=8.15^{+0.10}_{-0.12}$ keV, with equivalent width $63^{+54}_{-38}$ eV.  
The addition of this line to the model improves the fit by 
$\Delta\chi^2=41$. This excess is identified as a combination of the 
K$\beta$ line of Fe{\sc xxvi} (Ly$\beta$) and the 
K$\alpha$ line of Ni (Ni Ly$\alpha$; see discussion). 
The addition of this line-blend to our models 
does not change significantly the implied parameters of the 
rest of the fit. This line blend has a width consistent with that of the 
Fe K$\alpha$ line, although line widths should be revisited with the 
final {\it ASCA} calibration. 
Allowing an additional narrow line component 
within rest energy-range 6.4--7.0 keV did not 
improve upon the fit obtained with the Kerr model, 
$\Delta \chi^2 < 1$. The 90 \% upper limit on a narrow line 
with rest-energy between 6.4--7.0 keV is $\sim 80$ eV. 

In light of the evidence for ionized material, we also tested 
for the presence of an Fe K edge. The fit did not improve on addition 
of this model component ($\Delta \chi^2=0$). In order to make a 
direct comparison with Vaughan et al (1999) we calculated 
the 90 \% confidence upper limit 
on an edge at 8.76 keV (from He-like Fe), which  is $\tau < 0.09$. 
In light of the strong emission features it is difficult to 
understand the apparent absence of an Fe edge. We suspect the edge 
is present, and that the improved calibration of {\it ASCA} may later 
make it easier to precisely model this complex region of the spectrum. 

\subsection{X-ray and UV Absorption}

The spectral-energy-distribution of Akn~564 is relatively   
depressed in the optical and UV regime leading 
\citet{WalterFink93} to suggest  
that the unusual UV ($\lambda1375 $) 
to 2 keV flux ratio in Akn~564 could be due to absorption of 
the UV continuum. 
\citet{Crenshawea01} 
use the observed He{\sc ii} 
 $\lambda 1640/4686$ ratio and the 
continuum shape in the UV/optical regime to derive 
a  E(B-V) = 0.17 (Galactic =0.03, intrinsic=0.14) 
as the total reddening for Akn~564.  For a Galactic dust-to-gas ratio 
this implies a column $\sim 9 \times 10^{20}$\ cm$^{-2}$ 
of absorption in the X-ray band. We found this 
column of gas to be consistent with the X-ray data, 
and the presence of some absorber would 
lead to attenuation in the soft X-ray regime, and could thus 
explain the significant curvature (flattening) of the soft hump  
component  below 1 keV (\S~\ref{meanhump}). 
However, again we stress  that the
absolute form of the soft hump  is not the primary 
goal of this paper, rather the insights obtained from 
the variability of the component.

  \section{Spectral Variability\label{TimeSel}}	

	\subsection{Method and Selection Details} 
		\label{cutmethod}

\begin{figure}	
	\epsscale{0.95}
	\plotone{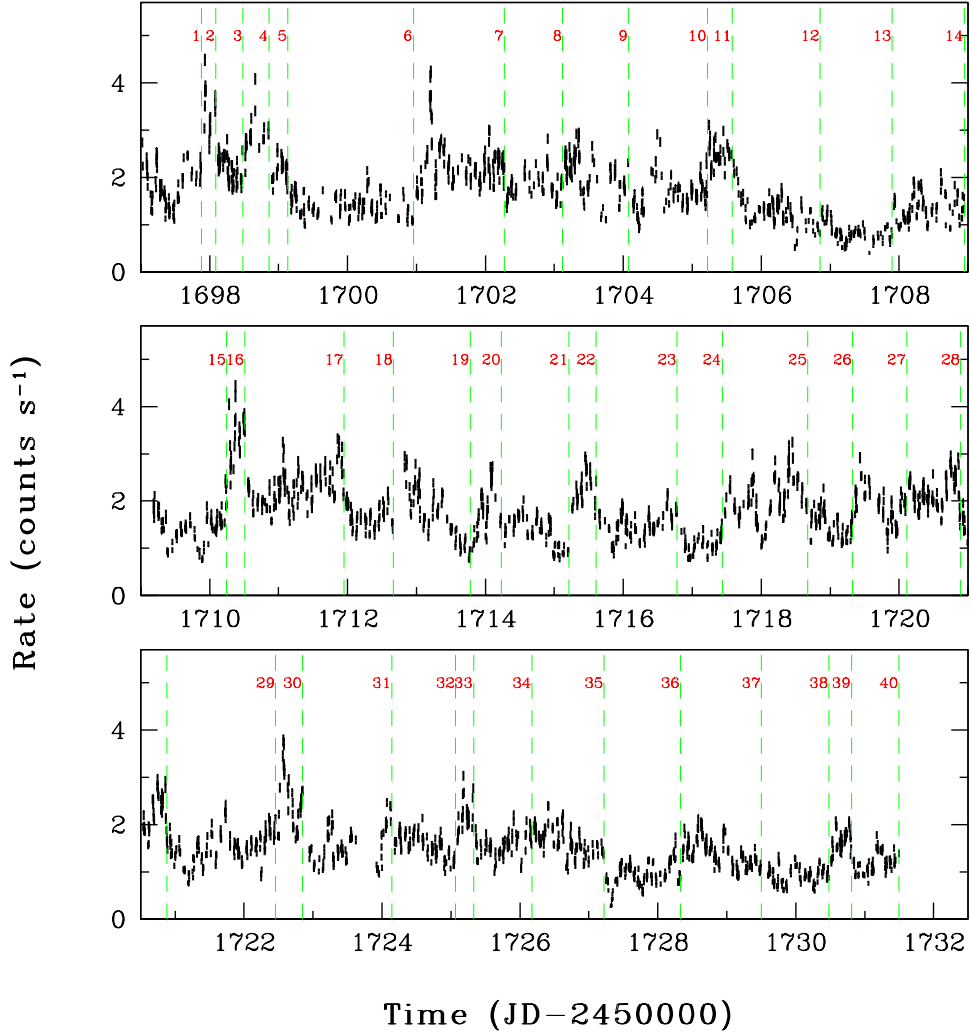}
	\caption[SIS Light curves]{Combined SIS 0.7--1.3 keV light 
curve in \ctssec{} and in 256 s bins. The background level in 
the source cell is about 4 \% of the source count rate, and not 
plotted. The vertical dashed lines delineate our 40 
time intervals within which spectra were extracted, as described  
in \S~\ref{TimeSel}.
\label{lcvcuts} } 
\end{figure}

In order to construct a complete picture of the multi-waveband 
variability of an AGN, one must consider the variations of X-ray spectral 
parameters. Simple flux-flux correlations can miss important clues to 
the emission and reprocessing mechanisms at work. 
In order to study the spectral evolution of Akn 564, we 
created 40 time-selected spectra across the 35 day observation. 
We sampled throughout the light curve following flares and dips 
using {\tt Xselect V2.0}. The resulting average baseline  for each 
time-selected spectrum was $\sim 75$ ks, and 
the average on-source exposure time was $\sim 25$ ks.  
(We note 
that the spectral variations which we will demonstrate and discuss 
were also evident when the data were sampled evenly, although 
such a sampling seemed to average out some interesting 
fluctuations). 
Figure~\ref{lcvcuts} shows our choice of intervals as vertical dashed 
lines plotted over the combined SIS light curve. 
We set the background, ancillary response, and response matrix files 
to be those of the mean spectrum, as 
the background spectrum and flux did not vary 
significantly during the 
observation, and we wished to attain the best possible signal-to-noise 
in the subtracted data. We also grouped the source counts with a 
minimum of 20 counts per energy bin as before. 
All fits were performed fixing the 
scaling factors for instrument normalization among the 
4 datasets (SIS and GIS) and the corrections for the SIS low-energy 
problem (see \S~\ref{SIStrouble}) to the best-fit values from 
the mean spectrum. We consistently modified our models with Galactic 
absorption by  a column $\NH = 6.4 \times 10^{20}$ cm$^{-2}$.
The time stamps of the light curves shown in this section are in 
JD-2450000 and refer to the mid-point of the observation.
Figure~\ref{40fits} shows the results from this analysis, described in
full below.

	\subsection{Variability of the Continuum\label{contfits}}

For each of the 40 time-selected spectra we fitted the 
SIS data in the range 2.5--4.88 keV simultaneously with 
the GIS data in the ranges 1.8--4.88 and 7.32--9.76 keV (all observers frame) 
using a simple power-law model (the same data 
exclusions as for the mean fit to the continuum slope, \S~\ref{meansp}).  
Figure~\ref{40fits} shows the light curves for the 
(model) continuum flux and the best-fit photon index $\Gamma$. 
Significant (see \S~\ref{CCresults}) but small 
variations are observed in $\Gamma$, which has a full 
range 2.45--2.72, (i.e.\ $\Delta\Gamma=0.27$) 
across the 35 days. Some significant changes are apparent down to timescales 
of $\sim 1$ day (Figure~\ref{40fits}). 
Fitting the photon indices to a constant model yields $\chi^2=67/39\ dof$. 
In addition to changes in slope, we note that the power-law component 
dominates the 2--10 keV band, thus rapid flux variations observed down 
to timescales of $\sim 1000$ s in that band  
are attributable to flux changes in the power-law continuum.

As an illustrative way of examining these spectral variations 
we constructed a plot to highlight the different aspects of the spectral 
evolution. 
Figure~\ref{allratios} shows ratio plots obtained by comparing the 
best-fit model for the first spectrum to the following 39 spectra. 
No fitting was performed. The plot is constructed to illustrate the 
variations of the spectrum and flux compared to the 
first day of data. 
Strong variations in both flux and continuum slope 
are present, as well as a strongly varying soft hump. 
We note that the soft hump is always evident 
above the power-law  continuum, although it appears to change 
in absolute strength and relative to the power law.  
The data show a hint of variations in the Fe K$\alpha$ line, which we 
investigate in more detail in \S~\ref{kacont}.  

	\subsection{Variability of the Soft X-ray Hump  
		\label{humpfits}}

To examine the variability of the soft hump  
we utilized SIS data in the range 0.75--1.7 and 2.5--4.88 keV 
simultaneously with GIS data in the range 1.0--4.88 and  
7.32--9.76 keV (observer's frame). The lower limit of the SIS was based 
on the level of agreement achieved between the two CCDs using our 
methods of correction, as described in \S~\ref{SIStrouble}. 
The model was a simple power-law plus a broad Gaussian 
for the soft hump, with Gaussian peak and width  fixed 
at the values noted in \S~\ref{meanhump}, while normalization of the 
soft hump, plus that of the continuum  power law  and the 
power-law slope were all left free. To avoid an overly complex 
model (which can result in false or local minima being found) 
we excluded data in the Fe K$\alpha$ regime (4.88-7.32 keV) 
for these fits. 
Figure~\ref{40fits} shows the time series for the normalization 
of the soft hump. The soft hump shows a marked 
decrease in flux across the 35-day observation.  
Fitting a constant model to the flux of the soft hump 
yields $\chi^2=870/39\ dof$. 
The hump shows a flux range 
of a factor $6.44\pm3.30$ while the 2--10 keV flux (when binned the same way) 
has a range of a factor $3.97\pm0.06$. 
An alternative measure to the trough-to-peak was obtained by averaging 
the first and 
last three of the 40 points for each component, this method avoids the 
numbers being dominated by a single strong event. This 
shows that the power-law falls by a factor $1.68\pm0.01$ and the soft hump by a 
factor $2.81\pm0.24$ from the start to the end of 
the \ASCA{} observation. The different amplitudes of  variability 
of the power-law and hump explain the gross  
change in softness ratio across the observation (Figure~\ref{lcv2}). 
By eye, it is clear that some of the flares in the hard X-ray flux 
are also evident in the soft hump. We discuss 
the detailed correlation between X-ray parameters in \S~\ref{CCresults}.

To confirm this result, we split the data into ``hard'' and ``soft'' states
based upon the softness ratio. 
We chose {\tt T} $< 10^{6}$ s for our soft-state spectrum, and 
{\tt T} $ > 2.23 \times 10^{6}$ s for our hard-state spectrum 
where {\tt T} is the time from the start of the observation 
(see Figure~\ref{lcv2}). Fits were performed on each state, in  
the same way as for the 40 individual spectra. Contour plots 
were generated for the normalization of the soft hump versus 
the photon index, $\Gamma$. Figure~\ref{softhard_cont} 
shows the contours for both states, confirming a significant 
decrease in the strength of the soft hump across the observation. 
The clear separation of contours illustrates that this is not 
confused with changes in the photon index, which are too small 
to explain the observed spectral changes anyway. 
$\Gamma$ 
is also plotted against the strength of the soft hump  
component, showing no evidence for a correlation between 
the two (Figure~\ref{nocorr}).  Another concern is that one 
might expect to see an 
anti-correlation between the fluxes of the hump and power-law 
if they were difficult to separate in the spectral fit. 
However, Figure~\ref{nocorr} demonstrates that the high-state data have a 
systematically higher strength for the soft hump than the low-state. This 
suggests that the two spectral components are well-separated in the fitting 
process, and that the two components have a close physical connection. 

Figure~\ref{allratios} indicates that the hump 
undergoes changes in shape on timescales down to days, 
 as evident (for example) by comparison of panels for time-cuts 13 and 35. 
However, the changes are small relative to the variations in 
flux of the hump, thus our model with fixed shape for the hump 
parameterizes most of the flux in the hump at each epoch. 
Thus, the evidence for some variation in hump shape does not 
compromise our approach in fixing 
the shape when testing for flux variability. 

We performed a final test, to check our result was not due to 
variations in the SIS soft response on timescales of weeks. 
While the SIS degradation is thought to be progressing slowly, the 
accelerated deterioration evident in data taken during calendar year 2000 
led us to look to the GIS for confirmation of the variability in the soft 
hump. Thus we excluded all SIS data below 1 keV, and used the GIS data 
down to 0.8 keV to determine the hump flux at each of the 40 epochs. 
This test strongly confirmed our result, that the soft hump flux falls 
during the course of this observation, and that this cannot be attributed 
to any instrumental effects in the {\it ASCA} data.

	\subsection{Variability of the Fe Emission Line}

\label{kacont}
In order to investigate the variability of 
the Fe emission line, we utilized SIS data in the range 2.5--7.32 keV 
simultaneously with the GIS data in the range 1.8--9.76 
keV. The model was a simple power-law plus the best-fitting 
profile to the Fe{\sc xxvi} K$\alpha$ line using the Kerr geometry, and the 
Gaussian model to the Fe K$\beta$/Ni K$\alpha$ blend, the latter  was linked 
to the flux of the K$\alpha$ line, using the observed ratio from the 
mean spectrum. 
The shape parameters of the Kerr (K$\alpha$) line were 
assumed to be constant. As a crude test of this assumption, we split the data 
using an intensity division equivalent to an SIS0 count rate 
of 2 \ctssec, that yielded high- and low-state spectra with similar 
signal-to-noise in each. 
We then fit for the mean continuum slope as described in \S~\ref{meansp}
and overlaid the data in the 4.88-7.32 keV band relative to the local 
power-law  slope in each case. The two line profiles are shown as 
data/model ratios in Figure~\ref{highlowfeprof}. The profiles appear 
similar, although there is some evidence for a slightly 
higher equivalent width in the 
low state. Using the high and low-state data we 
calculated contours of line 
flux versus $\Gamma$. Figure~\ref{highlowfecont} 
shows overlapping contours, thus this division 
of the data reveals no evidence for significant flux 
variability. 
We return to the variability of the line EW later. 

Returning to the 40 time-selected spectra, we fit those data 
with a model having free parameters of photon index, 
the normalization of the  power law, and the normalization of the 
line complex (using the shape parameters from the Kerr fit
to the mean profile), no model component was included for 
the soft hump, as the soft-band data were excluded for this test. 
Figure~\ref{40fits} shows the time series for 
the K$\alpha$ line. 
Fits to a constant model yield $\chi^2=36/39\ dof$. 
Thus the line does not show significant changes in flux when 
sampled on this timescale. 

As an alternative test we split the data into 8 intervals, to 
sample an intermediate timescale of several days per integration. 
The intervals each cover several of the original 40 time 
periods as follows: 1--6; 7--10; 11--17; 18--21; 22--26; 27--29; 
30--35; 36--40, and were chosen to follow apparent trends in line 
flux from Figure~\ref{40fits}. A model was constructed 
using the mean Fe line profile complex from \S~\ref{meanfek} 
and now assuming the line {\bf flux} also is constant. 
The soft-band data were excluded, as for all the Fe line
fits performed to date, and so the only free parameters in the 
fit were photon index and normalization of the power law. 
If the line is a constant flux component on top of the power-law 
throughout the observation, then by 
fitting for the local continuum slope and flux in each interval, 
one should be 
able to obtain an acceptable fit for each of the eight 
intervals. Figure~\ref{8fe} shows the data/model for each fit.
While some of the spectra show consistency with the mean line flux, 
several do not. For example, in interval 1, the line flux is 
significantly lower than the mean, while in interval 2, and 4,
it is higher. The shortest timescale on which one can see
some  indication for line variability is in comparison 
of the first two intervals (Figure~\ref{8fe}), taking the 
start time of the first interval, and stop time of the second then 
this    
constrains the line to arise in a region $< 700,000$ light seconds, 
or $\sim$ 1 light week in size. 
Taking the intervals with strongest contrast in residuals  
(1 and 4), we again calculated contours of line strength versus 
$\Gamma$. Figure~\ref{8fe_cont} shows these contours, which 
provide supporting evidence for line variability down to timescales of 
$\sim$ days. As an example, the difference in fit statistic 
for a line of fixed versus free strength for interval 2 is 
$\Delta \chi^2=14$, indicating the variability in line flux to be significant 
at $>$ 99 \% confidence. 

This result is supported by comparison of the Fe K$\alpha$ line in 
the soft-state versus hard-state data (defined in \S~\ref{humpfits}); 
a significant difference in line flux and EW is evident 
between those two states. 
For the soft-state we found EW$_{\rm soft}=491^{+103}_{-95}$ eV, with 
line normalization $n_{\rm soft}=6.2^{+1.5}_{-1.2} \times 10^{-5}$ 
\phflux; for the hard-state EW$_{\rm hard}=887^{+97}_{-114}$ eV, 
$n_{\rm hard}=9.6^{+1.1}_{-1.2} \times 10^{-5}$ \phflux. 
This variation cannot be an artifact of confusion with a 
change in photon index, as the soft and hard states are defined 
by the strength of the soft hump, and apparently unrelated to $\Gamma$. 
This result confirms that the line flux and equivalent width both vary during 
this \ASCA{} observation. 
There cannot be a simple correlation between soft hump 
flux and Fe line strength, if that was the case, we would expect to see 
a gradual decrease in line strength (in Figure~\ref{8fe}) as the 
hump strength went down dramatically across the observation; a 
correlated drop in line strength is not observed. 
We note the appearance of some absorption-type 
features at some epochs (e.g.\ interval 2), we do not investigate 
these apparent minor changes to line profile, as such small effects 
are best investigated when the final \ASCA{} calibration is 
made available.

   	\section{RMS Spectra} 
	\label{rms}

In optical/UV spectroscopy of AGN, 
it is easy to obtain a series of spectra of sufficiently high 
S/N to perform 
time-resolved spectroscopic analysis. In this case, 
a useful way to isolate variable features is to construct a 
root-mean-square (rms) spectrum. 
This ASCA long look has allowed us to use our 40 
individual time-selected spectra described in \S~\ref{TimeSel} 
in analogous way.  For each detector 
the 40 spectra have been rebinned so that each bin has at least a 5-sigma 
detection (but up to a maximum of 10 adjacent bins were combined 
to achieve a signal-to-noise ratio of 5 within a bin) 
then they were degraded to the 
resolution of the worst spectrum (i.e.\ the spectrum with worst S/N 
determined the actual bin widths for them all). 
We created mean and rms spectra by calculating the simple mean 
and rms flux in each bin. The choice of a simple mean as opposed to a 
weighted mean was dictated by the fact that we did not want to weigh 
in favor of high state spectra. 
 
The top panel in Figure~\ref{meanandrms} shows the
mean spectrum, the middle panel, the rms spectrum.
The bottom panel shows the  
rms spectrum after the power-law continuum is subtracted
(hence only the energy bands used in the fits are plotted). 
This shows that the most pronounced variability occurs in the soft-X band, 
consistent with our results of \S~\ref{sigmavar}, and that 
the variations in the Fe line are of low significance with this 
division of the data into 40 intervals.

   	\section{Cross-correlation Results
			\label{CCresults}}	

The X-ray light curves in different energy bands 
(Figure~\ref{lcv2}) exhibit  visually similar characteristics, 
suggesting a short time delay between the variations 
in each curve. In order to quantify any correlations, 
we undertook a cross-correlation analysis 
using the interpolation cross-correlation function (ICCF)
method of \citet[][]{GasSpar86} and \citet[][]{GasPet87} 
as implemented by \citet[][]{WP94}. 

We first considered a simple flux--flux correlation.  
We calculated the CCFs of the total hard-flux in the 2--10 keV band
with respect to the total soft-flux in the 0.7--1.3 keV band 
(binned to 5760 s). 
The maximum value of the correlation coefficient is $r_{\rm max}=0.942$ 
and the ICCF centroid $\tau_{\rm cent}=0.012^{+0.003}_{-0.011}$ d, less than 
0.02 d at 95 \% confidence. 
The CCF is sampled at a resolution of 0.005 day, and  
the centroids are computed using all points near the peak of the CCFs  
with values greater than 0.8 $r_{\rm max}$. The 1-$\sigma$ uncertainties 
quoted for $\tau_{\rm cent}$ are based on the model-independent 
Monte Carlo  method described by \citet{Petersonea98}. 
Given the $\sim$ 500 points in our light curves, there is a $\ll$ 0.1 \% 
chance of  exceeding $r_{\rm max} \approx 0.3$ from uncorrelated samples. 
The power-law component provides $\sim$ 75 \% of the flux in the 0.7--1.3 keV  
band (in the mean spectrum) and so rapid variability in the power-law  
flux is dominating the soft/hard flux correlation.  
To understand the physics, we also  
need to examine the spectroscopically-separated components.  
 
The cross-correlations between X-ray spectral parameters are   
summarized in Table~\ref{ccxx}. 
We calculated the CCFs of the light curves listed in Column (1) of
Table~\ref{ccxx} relative to the light curves shown in the first line, 
namely, the hard X-ray continuum, the soft X-ray hump, and the 
photon index $\Gamma$. 
The maximum value of the correlation coefficient $r_{\rm max}$
and the ICCF centroid $\tau_{\rm cent}$ relative to the hard X-ray 
continuum are given in Columns (2) and (3); the ones relative to 
the X-ray soft hump are given in Columns (4) and (5); and the 
ones relative to the photon index are given in Columns (6) and (7),
respectively. 
The CCFs are defined such that 
a positive lag means that the light curves listed in Line (1) of 
Table~\ref{ccxx} are leading the light curves listed in Column (1). 
The number of points in the X-ray light curves is 40; 
the number of points in the overlapping portion of the X-ray and 1365 
\AA{} light curves is 34. The CCFs are sampled at a resolution
 of 0.1 day, and the centroids are computed using all points near 
the peak of the CCFs with values greater than 0.8 $r_{\rm max}$. 
The 1-$\sigma$ uncertainties quoted for $\tau_{\rm cent}$ 
are based on the model-independent Monte Carlo method described 
by \citet{Petersonea98}. For our 40 points there is a 0.1 \% chance 
of exceeding $r_{\rm max} = 0.5$ from uncorrelated samples.
Figure~\ref{40fits} shows the light curves and the CCFs  
calculated relative to the hard X-ray continuum,  
and the hard X-ray continuum autocorrelation function (ACF). 
          
We see a strong correlation with no significant lag between the 2--10 keV flux 
and soft hump flux. The soft hump flux is distinct from the total soft-band
flux (which is the sum of power law plus soft hump counts), 
as it was obtained by spectral fitting. The use of quantities 
derived from spectral fitting limits the finest time resolution 
to $\sim$ a day. The correlation is good because events on timescales of 
$\sim $ a day occur in both the hump and power-law components. 
Also, while  the soft hump shows a $\sim$ 50 \% stronger decline over 
the 35 day observation than 
that which occurs in the 2--10 keV flux, both components show 
a gradual correlated decline. We also showed that the 
power-law flux varies down to timescales of $\sim 1000$ s and rapid 
changes in softness ratio \citep{Edelsonea01} indicate this occurs 
without a comparable change in the soft hump.   
Thus the strength of the correlation between the spectral components depends 
on the timescale of sampling. 

The final panel of Figure~\ref{40fits} 
compares the UV light curve at 1365 \AA\ with the 
hard X-ray flux. The correlation is strong, and the lag of the 
UV relative to the hard X-rays is 
consistent with $\sim$ 0, with an upper 
limit of 1.0 d at 95 \% confidence. 
The Fe K$\alpha$ line does not show significant correlations 
with any X-ray quantity (and thus is not shown in the Table or Figure) 
as the error on each of the 40 measurements is very large. 
It is also important to note that  $\Gamma$ is not well 
correlated with the UV flux 
or the soft hump flux (Table~\ref{ccxx}), ruling 
out simple models where a steepening 
of the photon index can explain observed properties in the UV 
and soft X-ray regimes. 

The strong correlation between the hard X-ray and UV curves leads 
us to make a direct comparison of the soft hump  flux and UV light curve. 
The CCF of 1365 \AA\  relative to the soft X-ray hump 
is shown in Figure~\ref{xscc}. 
There is a good correlation 
between soft hump and UV 1365 \AA\ flux, with the UV--soft hump 
lag consistent with 0 days and less than 1.0 d at 95 \% confidence.
Most notably, the bright X-ray flare around day 1710 is evident 
in the hard X-ray, soft hump and UV light curves. 
The difference between the correlation coefficients obtained 
for the hard-flux--1365 \AA\  versus hump--1365 \AA\ 
is not significant, so we cannot distinguish which is the 
primary link. 

	\section{Summary of Observational Results}
 
		\label{results}

\begin{itemize}
\item{\bf Akn~564 shows flux variations 
	by factors up to 16 in the 0.7--1.3 keV band, 
	14 in the 2--10 keV band when sampled on 256 s timescales 
	over a 35-day baseline.} 
\item{\bf Fractional variability amplitude 
	changes with time} with no clear correlations with flux or spectral 
	parameters. 
\item{\bf The mean photon index is $\Gamma \sim 2.54$ in the 
	hard band}, with variations of $\Delta\Gamma=0.27$ and 
	significant changes 
	observed down to a timescale of $\sim$ a day. 
\item{\bf A separate soft hump component detected below 1 keV 
	is found to be variable 
	down to timescales of $\sim$ a day}, 
	ranging by a factor of $\sim$ 6 in normalization 
	over the observation, but always being present. Some changes 
	in hump shape are evident down to timescales of days. 
\item{\bf The 
	photon index does not appear to be significantly 
	correlated with any of the other X-ray parameters} and 
	is not confused with soft hump strength. 
\item {\bf The powerlaw and hump are not well correlated on timescales 
	of 1000 s}
	Sharp variations occur in the power-law component, but examination 
	of the softness ratio during those times \citep{Edelsonea01}  
	indicates that comparable rapid changes do not occur in the 
	soft hump flux. 
\item {\bf The soft hump shows $\sim$ 50 \% larger variation in amplitude 
	than the power-law component over a baseline of weeks, when 
	sampled $\sim$ daily.} This difference causes changes in the 
	softness ratio across the observation.
\item{\bf The UV 1365 \AA\ flux is well correlated with the soft 
	hump and hard X-ray flux}. Correlated events (of different amplitudes) 
	appear in all three light curves on timescales of $\sim$ a day.  
	We detect no significant lag between the UV and X-ray bands. 
\item{\bf Fe K$\alpha$ emission is detected 
	in addition to a line representing a blend of Fe K$\beta$ emission 
	and Ni K$\alpha$, all from H-like  ions}, 
	confirming that the reprocessor is highly ionized.
\item{\bf The flux and EW of the Fe K$\alpha$ line  vary}, down to
	timescales of $\sim$ a week.
\end{itemize}

   \section{Discussion and Conclusions \label{discussion}}

Akn~564 shows large amplitude and rapid X-ray variability, with 
a trough-to-peak range of a factor of $\sim$ 16 sampled in the 
0.7--1.3 keV band using 256 s bins 
over this 35 day \ASCA{} observation. The 2--10 keV band also shows rapid 
 and large amplitude  variability with a maximum 
range a factor of 14 (using 256 s bins) and factors of $\sim$ several change 
over a few thousand seconds. 
While the fractional 
variability changes with time, this does not appear to 
be well correlated with flux. Furthermore, the 
fractional variability does not appear to be correlated with 
any X-ray, UV or optical parameter at zero lag 
\citep[see Paper III for the X-ray--optical correlations,][]{Shemmerea01}.
It is difficult to understand the apparent lag of the peak in 
flickering behavior  ($F_{\rm var}$ lags 
by 3 days relative to the big X-ray flares).
The BLS1 NGC~7469 also shows variations 
in $\sigma^2_{\rm rms}$ which do not appear to correlate with X-ray
spectral parameters of UV flux in any clear way. Akn~564 does not show 
the strong anti-correlation between fractional variability and UV flux 
indicated for NGC~7469 \citep{Nandraea2000}. 

\begin{figure}	
	\epsscale{0.85}
	\plotone{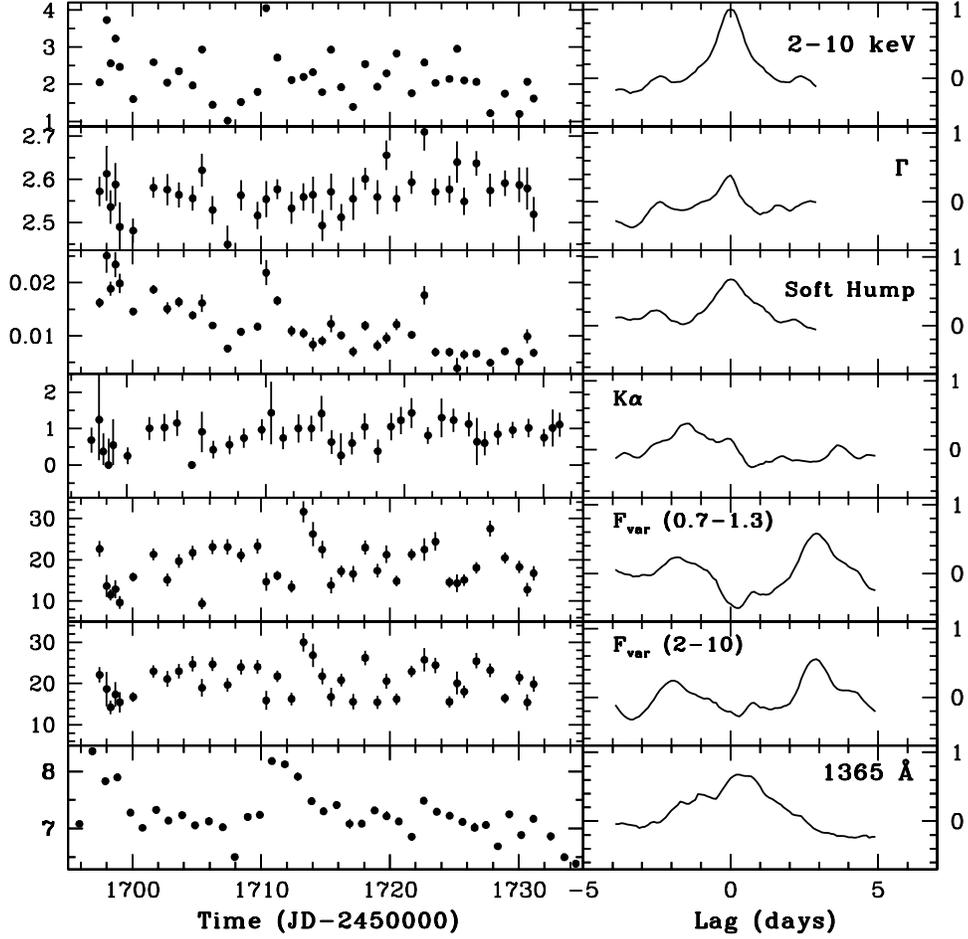}
	\caption[Time Series]{Spectral and timing parameters obtained 
from fits to the individual time-resolved spectra (left-hand column)
and  CCFs (right-hand column, discussed in \S~\ref{CCresults}). 
From the top, the light curves are 
the model continuum flux in the hard band, 
the photon index $\Gamma$, 
the soft hump normalization,
the $\Kalpha${} normalization, 
the fractional variability $F_{\rm var}$ in the soft and hard bands,
the continuum flux at 1365 \AA{}
from \citet{Collierea01}.
The X-ray continuum fluxes are in units of
$10^{-11}$ \eflux, the soft hump normalization and the 
$\Kalpha${} normalization are in units of $10^{-4}$ \phflux,
 the UV continuum in units of $10^{-15}$ \eflux{}.
The CCFs are calculated relative to the hard X-ray continuum  
(top, left), and the top panel on the right is the hard X-ray  
continuum ACF.
\label{40fits}}
\end{figure}

\begin{figure}	
	\epsscale{1.1}
	\plottwo{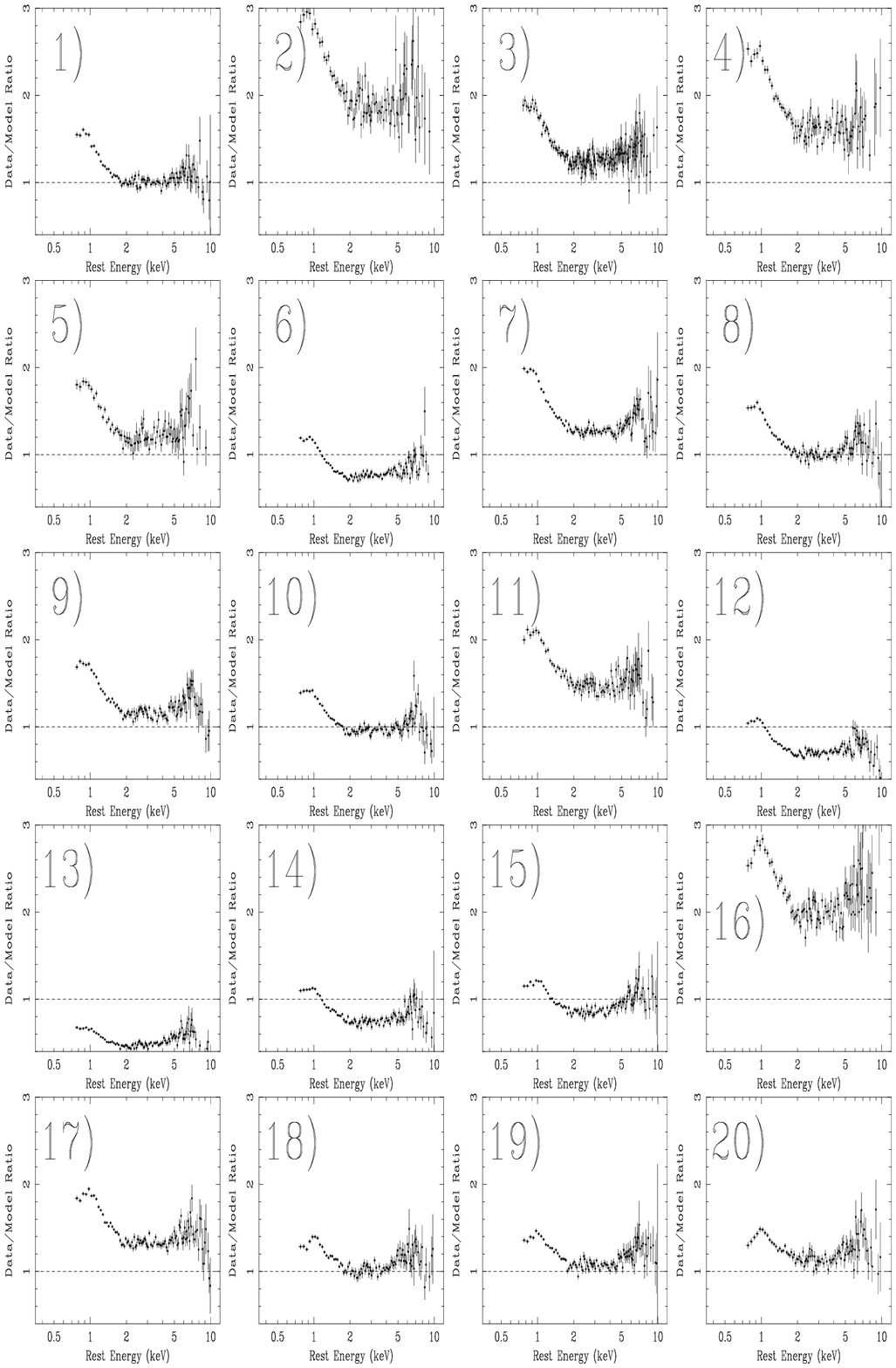}{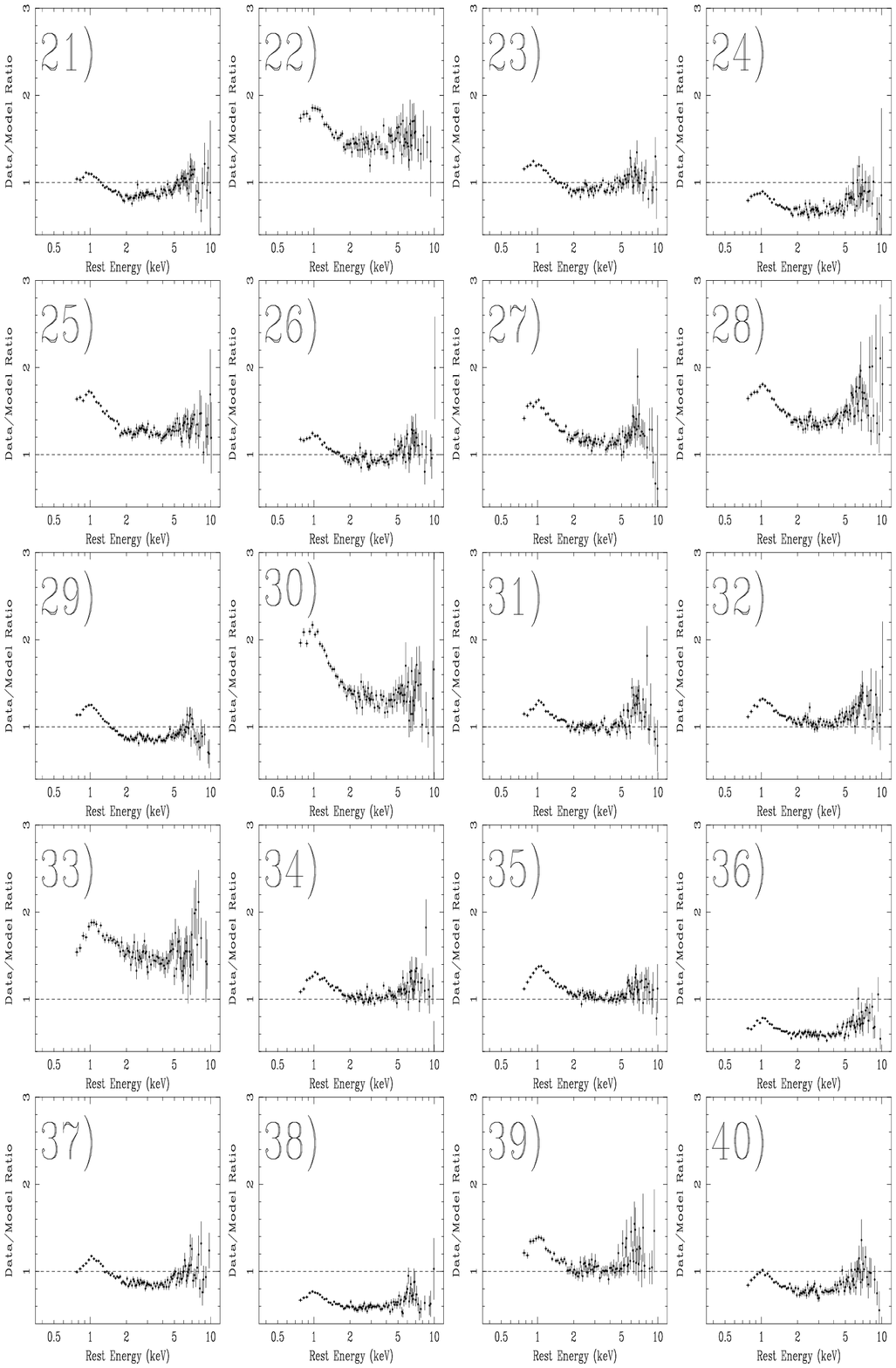}
	\caption[Ratio plots]{Ratio plots obtained by fitting the 
best-fit model for the first spectrum to the following 39 spectra.
Energy ranges and instrument utilized are described in 
\S~\ref{contfits}.
\label{allratios}}
\end{figure}

\begin{figure}	
	\epsscale{0.85}
	\plotone{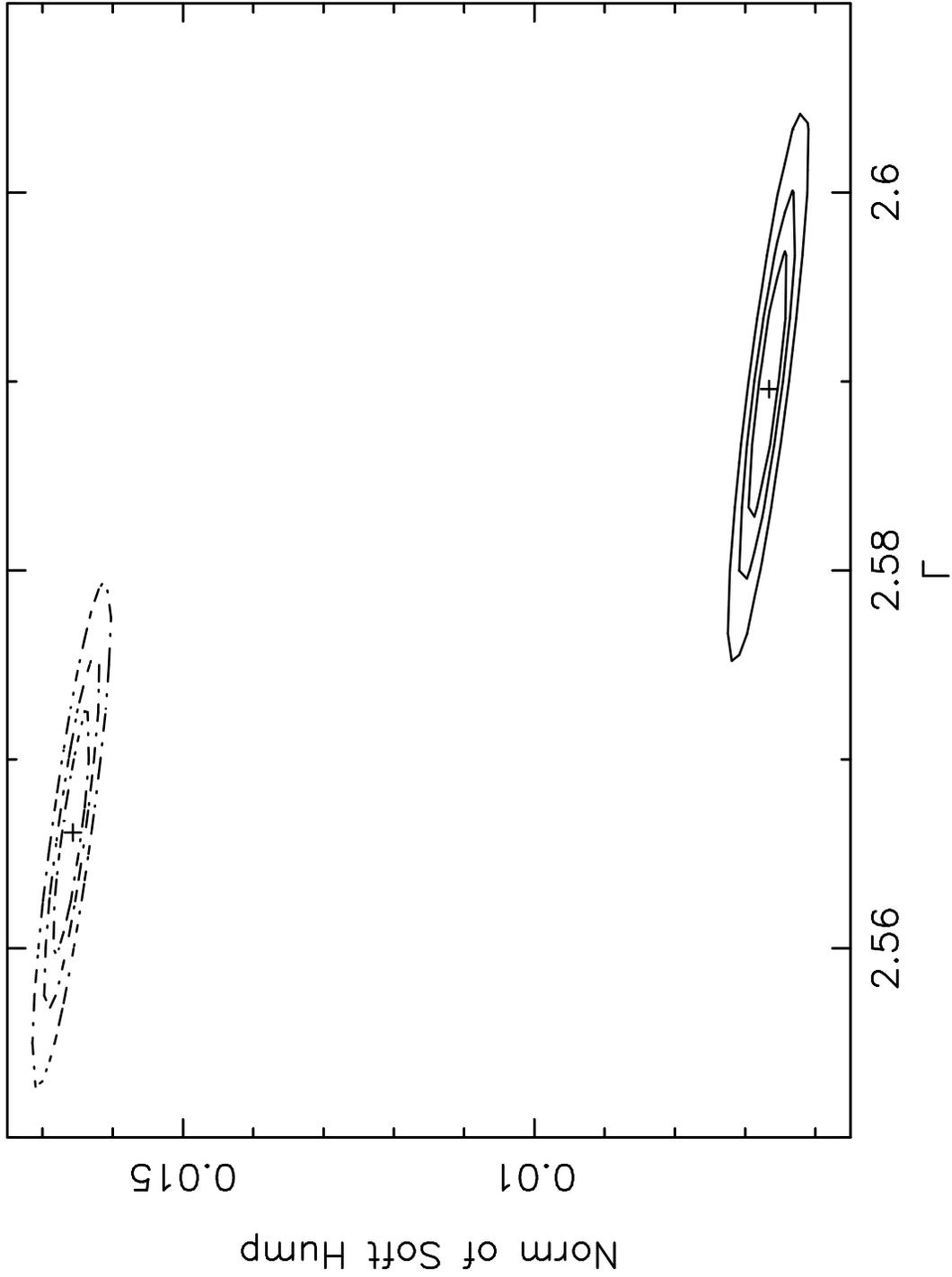}
	\caption[Soft Hump]{
The $\Delta \chi^2 = 2.3, 4.61, 9.21$ contour levels for 
the soft hump normalization (in units \phflux) 
vs.\ photon index $\Gamma$.
The dashed contours correspond to the soft state, 
the full contours to the hard state, and 
the best-fit values are indicated by crosses.
\label{softhard_cont}}
\end{figure}

\begin{figure}	
	\epsscale{0.85}
	\plotone{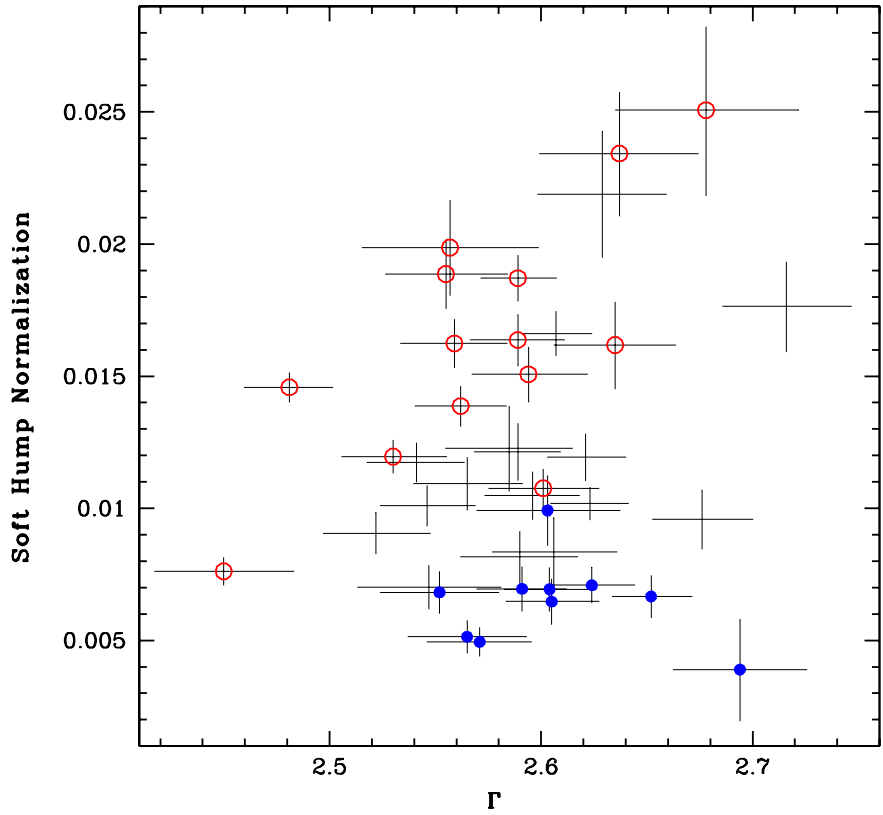}
	\caption[Soft Hump versus Photon Index]{
The strength of the soft hump plotted against 
photon index $\Gamma$, with high-state 
(open circles) and low states (filled circles) overlaid.
\label{nocorr}}
\end{figure}

\begin{figure}	
	\epsscale{0.95}
	\plotone{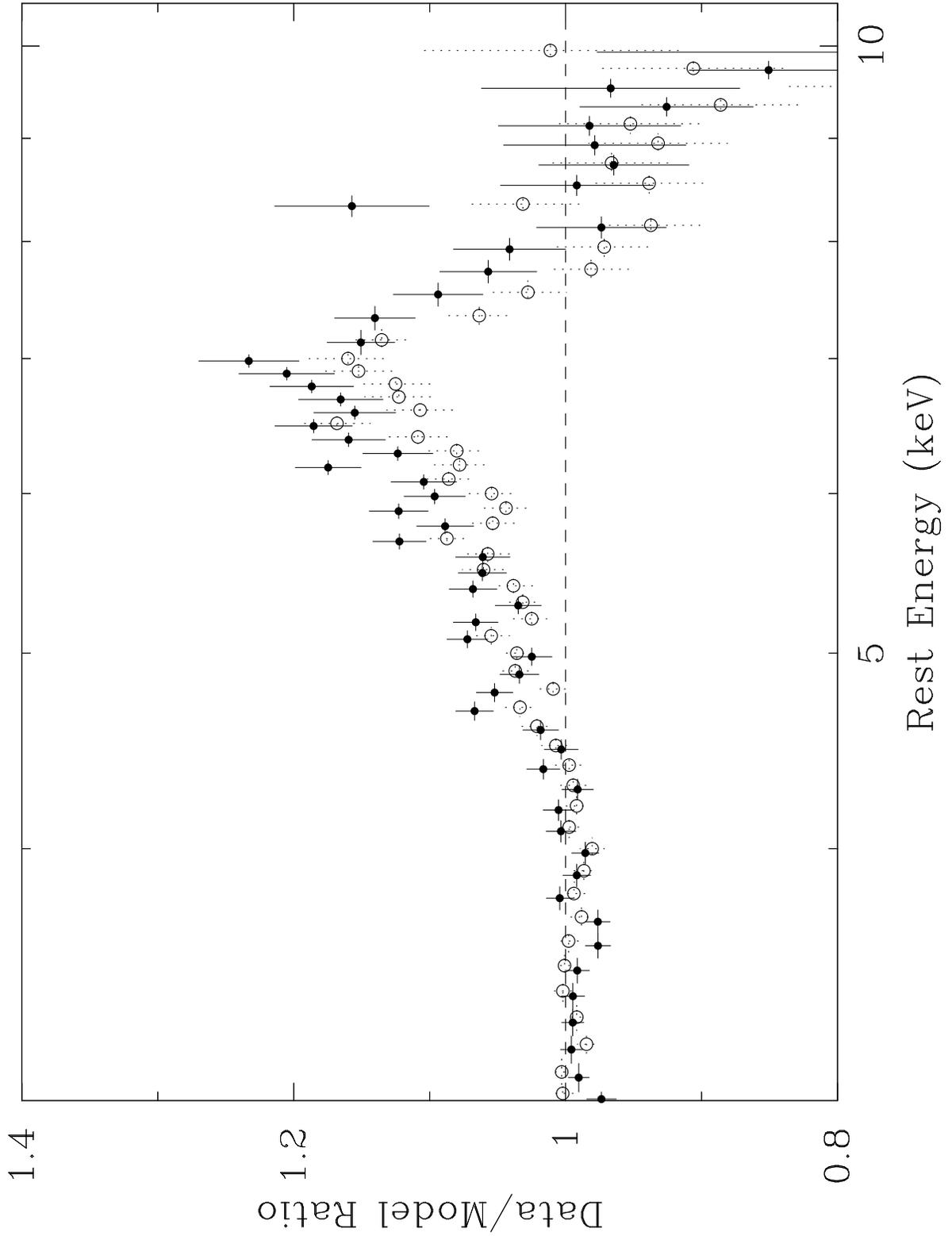}
	\caption[High$/$Low Fe profile]{Data from the Fe K 
regime compared to the continuum model, with high-state 
(open circles) and low states (filled circles) overlaid.
\label{highlowfeprof}}
\end{figure}

\begin{figure}	
	\epsscale{0.85}
	\plotone{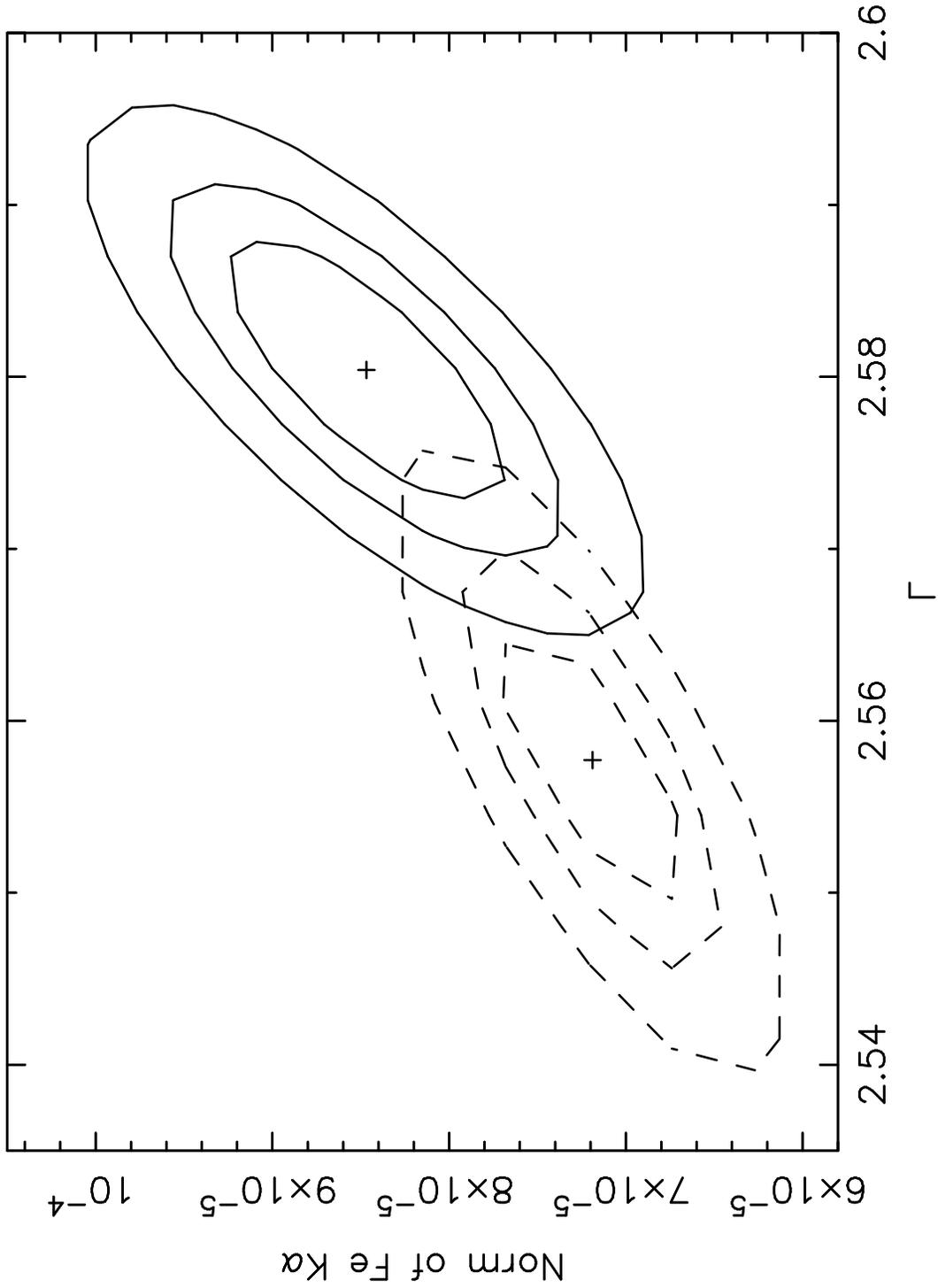}
	\caption[High-Low contours of Fe K$\alpha$]{
The $\Delta \chi^2 = 2.3, 4.61, 9.21$ contour levels for 
Fe K-shell line intensity (in units \phflux) 
vs.\ photon index $\Gamma$. 
The dashed contours correspond to the low state, 
the full contours to the high state, and 
the best-fit values are indicated by crosses.
\label{highlowfecont}}
\end{figure}

\begin{figure}	
	\epsscale{0.95}
	\plotone{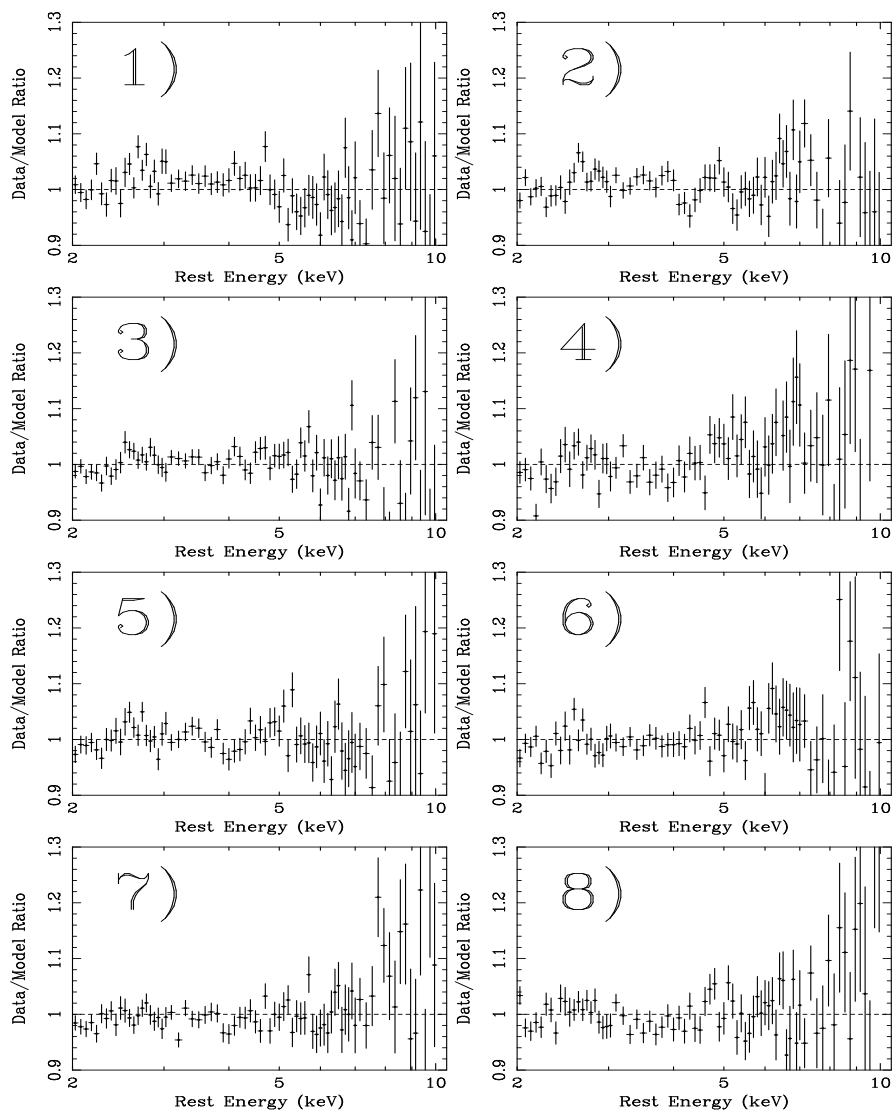}
	\caption[Time Variable Fe profiles]{Data from the Fe K 
regime compared to the continuum model plus fixed 
line profile from the mean spectrum, sampled every few days.  The 
systematically high or low residuals indicate line variability, 
see \S~\ref{kacont} for more details.
\label{8fe}}
\end{figure}

\begin{figure}	
	\epsscale{0.85}
	\plotone{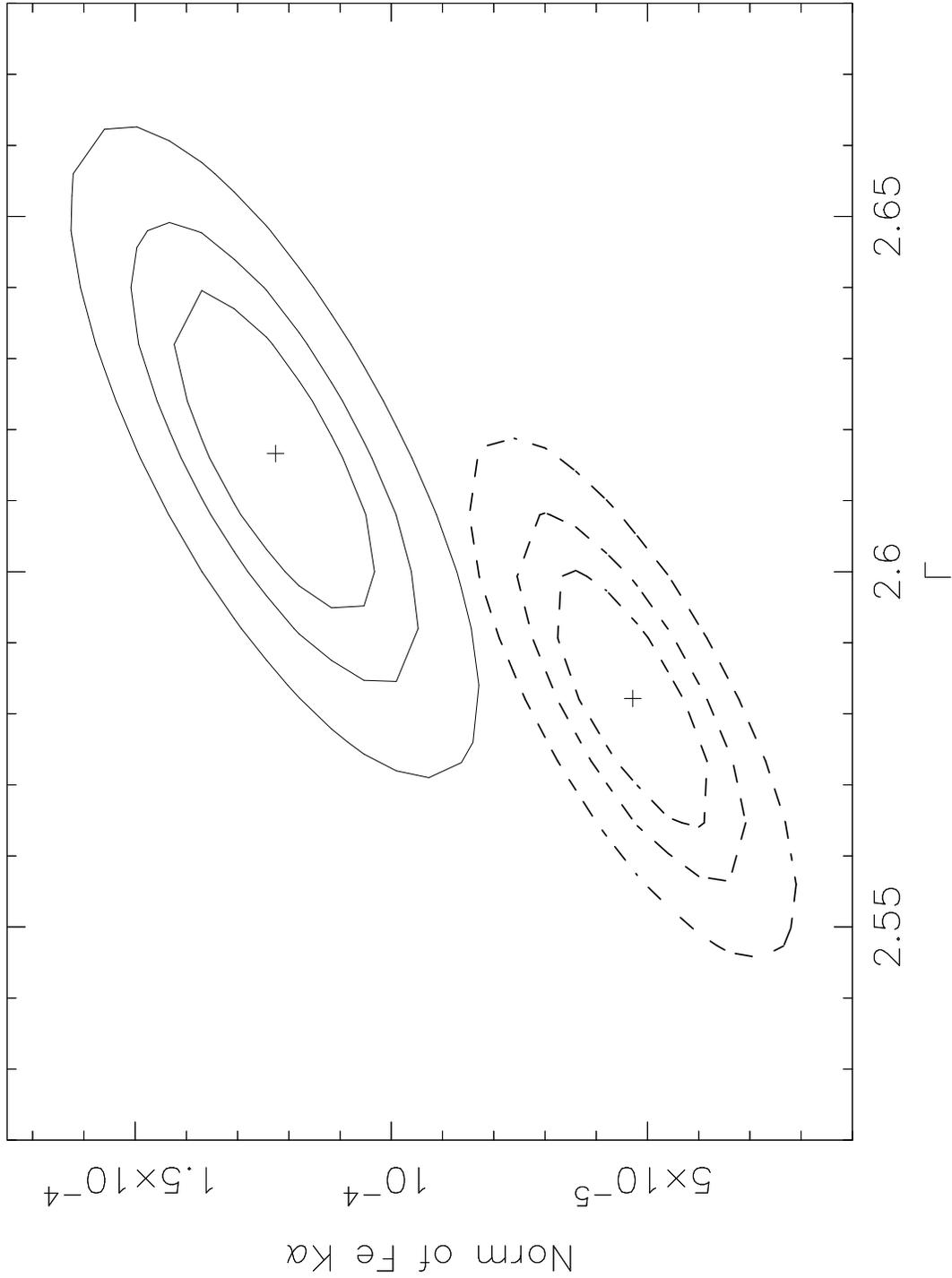}
	\caption[High-Low contours of Fe K$\alpha$]{
The $\Delta \chi^2 = 2.3, 4.61, 9.21$ contour levels for 
Fe K shell line intensity (in units \phflux) 
vs.\ photon index $\Gamma$. 
The dashed contours correspond to interval 1 
the full contours to interval 4, as 
described in \S~\ref{kacont}.
\label{8fe_cont}}
\end{figure}

\begin{figure}  
        \epsscale{0.95}
        \plotone{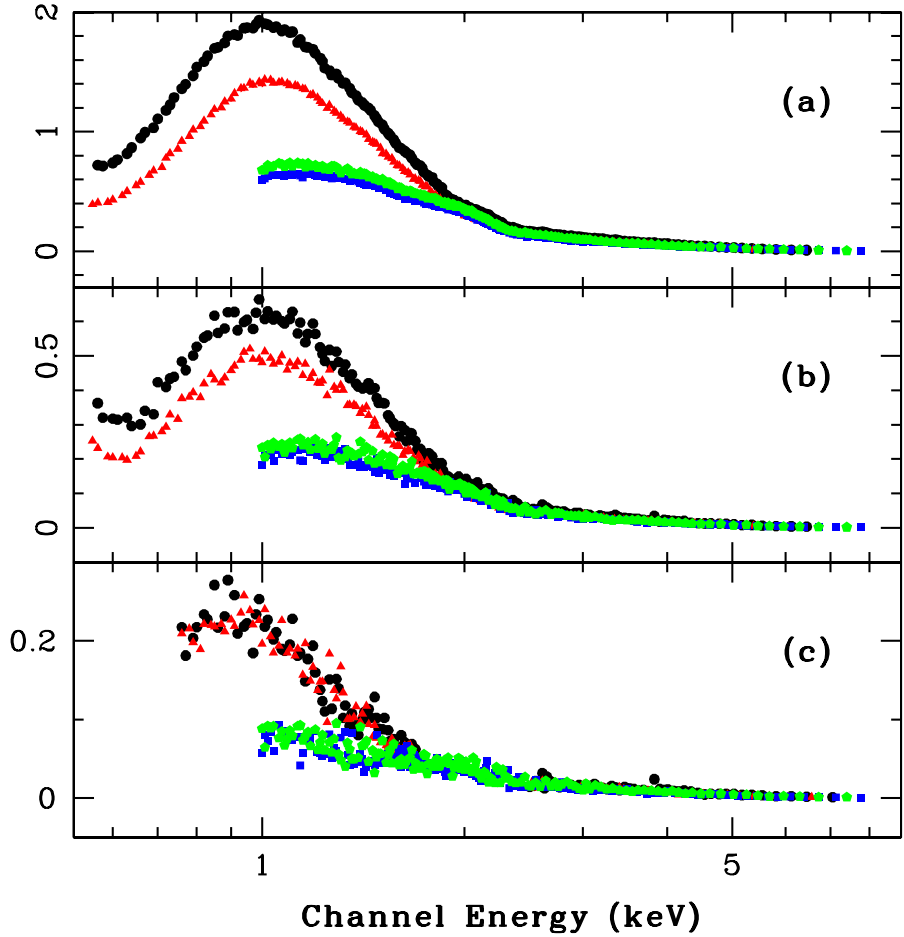}
        \caption[Mean and RMS spectrum]{The top panel shows the
mean spectrum obtained from our 40 time-selected spectra 
(in units of counts s$^{-1}$ keV$^{-1}$, errorbars in the vertical  
direction are included);
the middle panel shows the rms spectrum, that isolates the 
variable parts of the spectrum; the bottom panel shows the 
rms spectrum after the power-law continuum is subtracted.
Circles refer to the  SIS-0 spectrum; triangles  to the SIS-1; 
squares to the GIS-2;
pentagons to the GIS-3. 
\label{meanandrms}}
\end{figure}

\begin{figure}	
	\epsscale{0.90}
	\plotone{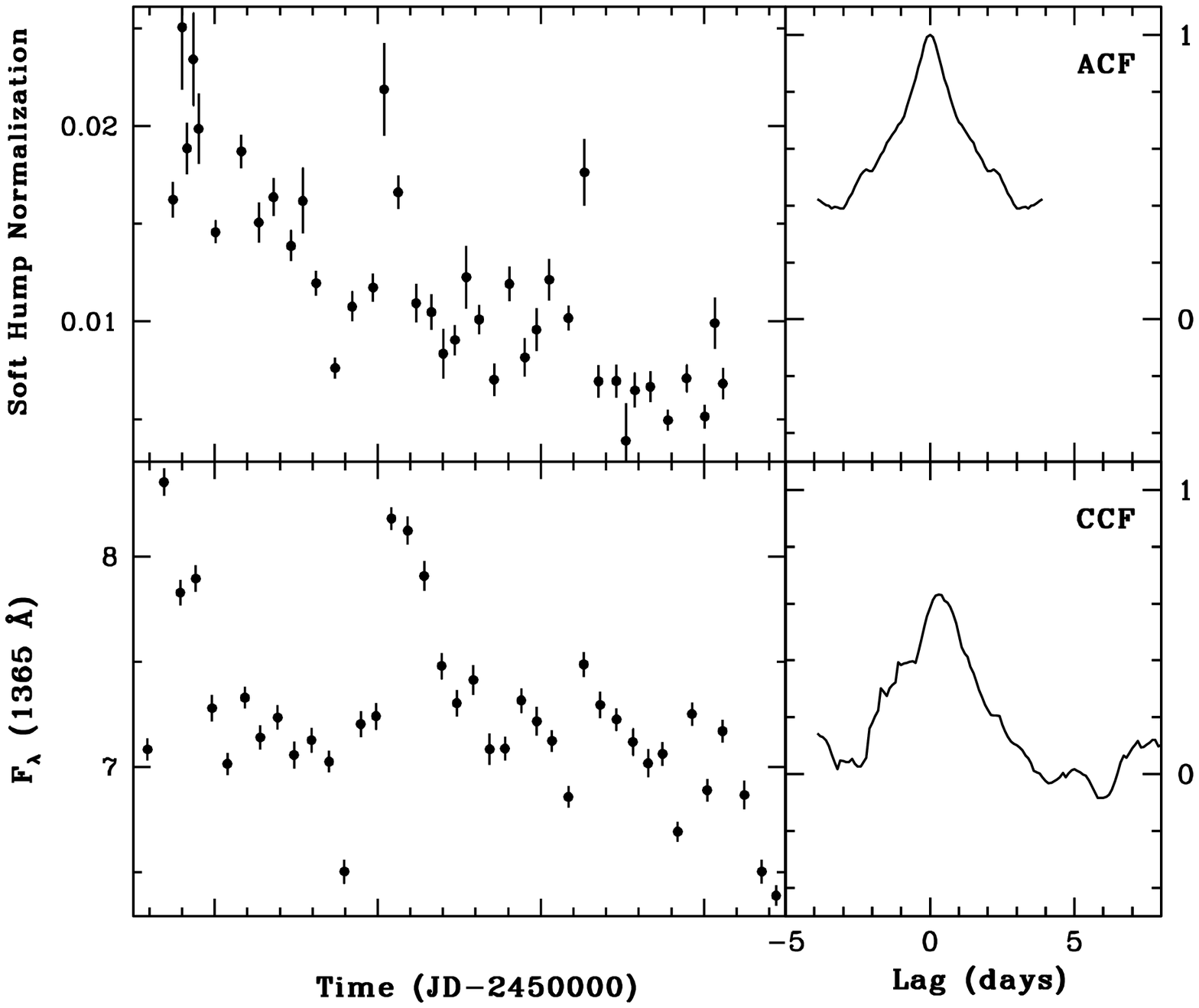}
	\caption[Light curves (left-hand column) and CCFs 
(right-hand column).]{
Light curves (left-hand column) and CCFs (right-hand column). 
From the top, the light curves are the X-ray soft hump 
normalization, and the UV continuum at 1365 \AA. 
The X-ray soft hump normalization is in units of  
$\phflux$, and the UV fluxes in units of $10^{-15}$  \efluxA. 
The CCF is calculated relative to the X-ray soft hump 
(top, left), and the top panel on the right is the X-ray 
soft hump ACF. 
Positive peaks mean that the X-ray soft hump is leading.
UV data from \citet[][]{Collierea01}.
\label{xscc}}
\end{figure}

The photon index $\Gamma \sim 2.54$ is steep 
for a NLS1 galaxy compared to measurements of other such sources 
\citep{BME97,TGN98,Leighly99II,Vaughan99b}. The 
photon index varies between 2.45--2.72, comparable 
to the variability in $\Gamma$ observed in the BLS1 NGC~7469 
\citep[$\Delta\Gamma=0.32$,][]{Nandraea2000}. This may indicate 
that the disk corona shows a similar degree 
of fluctuation in temperature and/or  
optical depth in BLS1s and NLS1, although the absolute conditions 
appear to be very different. 
In contrast to NGC~7469, there 
appears to be no correlation between photon index and UV flux 
(Figure~\ref{40fits}). In addition, the extrapolation of the hard X-ray
power law  overpredicts the UV and optical fluxes \citep{WalterFink93}.
Thus the hard X-ray power law does not appear to simply 
extrapolate into the UV band in Akn~564, and the component must 
terminate or turn over at wavelengths shorter than 1365 \AA.

A distinct soft hump is observed in addition to the power-law continuum.
The component rises above the power-law below 2 keV, and shows 
a distinct spectral flattening below 1 keV, rising again below 0.5 keV.  
This hump was first detected in an earlier \ASCA{} observation of 
Akn~564 \citep{TGN99}. The soft hump 
shows flares which are correlated with those in the power-law component, 
down to timescales of days (the shortest timescale we can reliably probe 
the hump spectroscopically using these data), the flares are 
not generally of the same amplitude in the soft and hard bands. 
There is a  distinction in the soft hump and power-law 
variability over a timescale of  weeks, with the former varying by 
a factor of $\sim 6$ across the 35-day baseline of the observation compared 
to a factor 4 in the power-law (when 
sampled $\sim$ daily). This difference in amplitudes of variation 
causes changes in the softness ratio across the observation. 

We can infer something about the 
rapid variability of the hump only by using an indirect method. 
Some spectral changes on timescales of $\sim$ 1000 s \citep{Edelsonea01} 
are clearly due to sharper changes in hard than soft X-ray flux 
suggesting the soft hump does not vary as 
fast as the power-law. 
In this case, the overall larger measure of $F_{\rm var}$ in the  
soft band is most likely due to 
the relatively strong changes in hump-flux on timescales of weeks. 

 We note that the pronounced variability of the soft hump 
immediately rules out an origin as starburst emission.
{\it Chandra} results from another NLS1, TonS180 \citep{Turnerea01}, have 
shown no absorption lines, indicating the shape of the hump in that case 
is unlikely to be due to a combination of absorption effects and 
a steepening continuum form. Akn~564 is more likely to show 
a soft component which is modified in shape by 
absorption, as the UV data indicate the presence of a column 
$\sim 9 \times 10^{20}{\rm cm^{-2}}$ \citep{Crenshawea01}. 
However, the fact that the {\it ASCA} 
spectra of Akn~564 and TonS180 look very similar has 
led us to consider the soft hump in Akn~564 likely to be 
a continuum component or broadened reprocessed component 
from the accretion disk, as suggested for TonS180. 
\citet{Collea01} find a soft hump in NGC~4051 of similar form 
to that in Akn~564 and TonS180 and this may also be 
attributable to a smooth continuum-like component. 
This good correlation between the UV flux and the soft hump 
and the upper limit to the lag of $\sim 1 $ d 
indicates a contribution to both bands by the same physical 
component, perhaps emission from the accretion disk. 
The UV/soft hump correlation  and the apparent association of the soft hump 
with NLS1s appears consistent 
with the idea that the disk peaks at a higher temperature in NLS1s than 
BLS1s, as expected if we are viewing thermal emission 
from a disk which is 
more highly ionized in the NLS1 case \citep{RFM92}. Unfortunately these 
data do not allow 
us to distinguish whether the soft hump is due to 
reflected radiation from an ionized 
disk, thermal emission from the disk or a combination of both. 
The changes in 
the shape of the hump may indicate that it is the superposition of 
several different components of emission/reflection, which 
possess differing timescales of variability. 

In general, we find disk-corona models to be challenged 
by the lack of correlation of $\Gamma$ with the strength of the 
soft hump, presumably the seed photon source, which
should cool the corona and cause a steepening in the power law 
as it increases in strength. One possibility is that 
the corona cooling is saturated, i.e. that even a significant 
reduction in 
the soft hump leaves a large enough soft-photon reservoir that 
the resulting spectrum is always strongly cooled. 
We defer a more detailed discussion of the multiwaveband 
lags and correlations, and reprocessing models, to Paper III 
\citep{Shemmerea01}. 

An Fe K$\alpha$ line of high equivalent width 
is detected and clearly attributable to highly ionized material. 
These are strong observational results which 
are robust to the current inaccuracies in calibration. 
Analysis using the current calibration suggests 
a dominance of Fe{\sc xxvi} ions 
in Akn~564 (yielding Ly$\alpha$), although 
a blend of emission from several states is most likely. 
An additional line  
of equivalent width  $63^{+54}_{-38}$ eV is  detected,  
close to a rest energy of 8.2 keV. Lines consistent with 
this energy are the K$\beta$ line from H-like Fe (Ly$\beta$) 
and the K$\alpha$ line 
from H-like Ni (Ly$\alpha$).   
Assuming solar abundances then the predicted equivalent 
width of Fe K$\beta$ is 77 eV, and that of Ni K$\alpha$ is 
35 eV, based on the strength of the Fe K$\alpha$ line. 
Thus we attribute the line at 8.2 keV to a blend of these two, 
and the total observed equivalent width is consistent with  
such a sum. A previous short \ASCA{} observation 
of Akn~564 showed ambiguity 
between ionization-state of the reprocessor and inclination 
angle of the disk system \citep{TGN99}, however, these data 
show that an ionized reprocessor is clearly present in this source. 
The discovery of ionized reprocessors in some 
BLS1s \citep[e.g.,][]{Guainazziea98} indicates that the luminosity of the 
central source may be as important as accretion rate in creating 
an ionized surface to the disk. 

The equivalent widths for the Fe lines are huge (with 
EW$=653\pm85$ eV  for Ly$\alpha$) compared to those 
observed in Seyfert~1 galaxies \citep[e.g.][]{Nandraea97a}.  
Similarly large values 
have been noted in previous \ASCA{} observations of this and other NLS1s 
\citep{TGN98, Comastriea98}. 
The large EWs may be partly explained by the relatively large 
line fluxes expected from He-like and H-like ions 
compared to neutral material. Another possibility is that 
NLS1s have high Fe abundance. Noting the high EWs of Fe K$\alpha$
and the strong Fe{\sc ii} emission in NLS1s 
\citet{TGN99} suggested that high Fe abundance 
might be a general property of NLS1s;  
furthermore, high metallicities are expected in some evolutionary 
models for 
NLS1s \citep{math2000}. 
The high EW cannot be due to a large solid angle for the reprocessor or 
a hidden hard X-ray component, as, in either of these cases we 
would observe a detectable flattening to high energies due to
the presence of the associated Compton hump. 
The \ASCA{} data show variations in the Fe K$\alpha$ 
flux by a factor $\sim 2.4$ 
on timescales $< 7 \times 10^5$ light seconds, indicating the 
bulk of the iron line originates within $\sim $ a light week of the nucleus.

\acknowledgements

TJT is pleased to acknowledge support for this work by NASA through 
grant number NAG5-7385 (LTSA). We also acknowledge support from 
HST--GO--08265.01--A from the Space Telescope Institute, 
which is operated by the Association of Universities for Research in 
Astronomy, Inc., under NASA contract NSS5-226555. 
SM acknowledges support through NASA grant NAG5-8913 (LTSA).
We thank Ken Pounds, Steve Kraemer, Mike Crenshaw,  Alex Markowitz, 
Hagai Netzer, Ohad Shemmer and the anonymous referee for useful comments. 
We are grateful to the \ASCA{} team for their operation of the 
satellite, and to Tahir Yaqoob for discussions on the \ASCA{} 
calibration.

\clearpage 

\clearpage

\clearpage 

\clearpage

\clearpage 

\clearpage

\clearpage 

\clearpage

\clearpage

\clearpage 

\clearpage 

\clearpage 

\clearpage

\clearpage

\clearpage 

\begin{deluxetable}{lcccccc}
\tablewidth{0pc}
 \tablecaption{X-Ray and UV Cross-Correlation Results\label{ccxx}}
\tablehead{
\colhead{} & \multicolumn{2}{c}{F$_{2-10 \;\,{ } {\rm  keV}}$} &
\multicolumn{2}{c}{Soft Hump}
& \multicolumn{2}{c}{$\Gamma$} \\
\colhead{} & \colhead{$r_{\rm max}$} & \colhead{$\tau_{\rm cent}$\tablenotemark{a}} 
& \colhead{$r_{\rm max}$} & 
\colhead{$\tau_{\rm cent}$\tablenotemark{a}} &  \colhead{$r_{\rm max}$} 
& \colhead{$\tau_{\rm cent}$\tablenotemark{a}}\\
\colhead{(1)} & \colhead{(2)} & \colhead{(3)} & 
\colhead{(4)} & \colhead{(5)} & \colhead{(6)} & \colhead{(7)}
}
\startdata
$\Gamma$                  & 0.389 & -0.05$^{+0.10}_{-0.35}$ & 0.058 &
0.00$^{+0.04}_{-0.59}$ & \nodata & \nodata  \\
Soft Hump               & 0.675 &  0.00$^{+0.20}_{-0.10}$ & 
\nodata\tablenotemark{b} & \nodata\tablenotemark{b}  & \nodata & \nodata \\
$F_{\rm var}$ (0.7--1.3 keV)   & 0.584 & $2.94^{+0.31}_{-0.15}$  &  0.232  &
3.00$^{+1.10}_{-0.15}$ & 0.153 & -0.95$^{+2.25}_{-0.84}$   \\              
$F_{\rm var}$ (2--10 keV)      & 0.557 & $2.85^{+0.30}_{-0.10}$  &  0.405  &
3.05$^{+2.26}_{-0.10}$ & 0.239 & -0.00$^{+3.89}_{-5.54}$  \\    
F$_{\lambda}$ (1365 \mbox{\AA}) & 0.683 & 0.39$^{+0.26}_{-0.84}$ & 
0.633 & 0.35$^{+0.35}_{-0.44}$ & 0.112 & $0.70^{+0.30}_{-0.30}$ \\
\enddata
\tablenotetext{a}{1-$\sigma$ uncertainties.}
\tablenotetext{b}{The Soft Hump ACF is shown in Figure~\ref{xscc}.} 
\end{deluxetable}
\end{document}